\documentclass[lettersize,journal]{IEEEtran}
\IEEEoverridecommandlockouts
\usepackage{cite}
\usepackage{amsthm}
\usepackage{amsmath,amssymb,amsfonts}
\usepackage{algorithm}
\usepackage{algorithmic}
\usepackage[super]{nth}

\usepackage{graphicx}
\usepackage{textcomp}
\usepackage{xcolor}
\usepackage{bm}
\usepackage{graphicx,epstopdf,psfrag,url,cite,xcolor,multirow,float}
\usepackage{color}
\usepackage{array}
\usepackage{subfigure}
\usepackage{graphicx}   

\graphicspath{{figures/}}
\def\BibTeX{{\rm B\kern-.05em{\sc i\kern-.025em b}\kern-.08em
    T\kern-.1667em\lower.7ex\hbox{E}\kern-.125emX}}
\begin{document}
\title{Spatio-Temporal Attention Enhanced Multi-Agent DRL for UAV-Assisted Wireless Networks with Limited Communications}



\author{
Che Chen, Lanhua Li, Shimin Gong, Yu Zhao, Yuming Fang, Dusit Niyato\\

\thanks{
Che Chen, Lanhua Li, and Shimin Gong are with the School of Intelligent Systems Engineering, Shenzhen Campus of Sun Yat-sen University, Shenzhen 518000, China, and Southern Marine Science and Engineering Guangdong Laboratory (Zhuhai), Zhuhai 519082, China (e-mail: chench576@mail2.sysu.edu.cn, \{lilh65, gongshm5\}@mail.sysu.edu.cn).

Yu Zhao is with the Department of Equipment Management and Unmanned Aerial Vehicle Engineering, Air Force Engineering University, Xi'an (e-mail: zhaoyuair@163.com).

Yuming Fang is with the School of Computing and Artificial Intelligence, Jiangxi University of Finance and Economics, Nanchang 330032, China (e-mail: fa0001ng@e.ntu.edu.sg).

Dusit Niyato is with the College of Computing and Data Science, Nanyang Technological University, Singapore (e-mail: dniyato@ntu.edu.sg).
}

\vspace{-0.3cm}
}

\maketitle
\thispagestyle{empty}

\begin{abstract}
In this paper, we employ multiple UAVs to accelerate data transmissions from ground users (GUs) to a remote base station (BS) via the UAVs' relay communications. The UAVs' intermittent information exchanges typically result in delays in acquiring the complete system state and hinder their effective collaboration. To maximize the overall throughput, we first propose a delay-tolerant multi-agent deep reinforcement learning (MADRL) algorithm that integrates a delay-penalized reward to encourage information sharing among UAVs, while jointly optimizing the UAVs' trajectory planning, network formation, and transmission control strategies. Additionally, considering information loss due to unreliable channel conditions, we further propose a spatio-temporal attention based prediction approach to recover the lost information and enhance each UAV's awareness of the network state. These two designs are envisioned to enhance the network capacity in UAV-assisted wireless networks with limited communications. The simulation results reveal that our new approach achieves over 50\% reduction in information delay and 75\% throughput gain compared to the conventional MADRL. Interestingly, it is shown that improving the UAVs' information sharing will not sacrifice the network capacity. Instead, it significantly improves the learning performance and throughput simultaneously. It is also effective in reducing the need for UAVs' information exchange and thus fostering practical deployment of MADRL in UAV-assisted wireless networks.
\end{abstract}



\section{Introduction}
Recently, the applications of unmanned aerial vehicles (UAVs) have attracted extensive attention in various wireless networks, such as UAV-assisted sensing networks, mobile edge computing, and wireless-powered networks~\cite{2023AssistedDisasterRescue,2023UAVHosseinMotlagh,basharat2022resource}. UAVs offer high mobility, adaptability, and controllability, which improve the efficiency of data transmission in these networks. Specifically, UAVs can dynamically relocate to enhance channel conditions for data transmissions from ground users (GUs) to the base station (BS), improving both network coverage and capacity. The UAV can also serve as a mobile energy supplier for low-power GUs in large-scale wireless-powered networks to prolong the network lifespan~\cite{2022WPT}. However, each UAV faces limitations in energy supply, computation capacity, and coverage area. These limitations can be practically alleviated by employing multiple UAVs to serve the GUs in large-scale wireless networks~\cite{2023UAVSurveyJav,2022SecureEnergy-Efficient}, by designing efficient information sharing and collaborative control mechanisms for UAVs' trajectory planning, multi-hop networking, and transmission control strategies.

\vspace{-0.2cm}
\subsection{Challenges and Motivations}
The UAVs' mobility control can be achieved by joint trajectory planning, allowing UAVs to simultaneously serve different GUs to improve the network coverage and transmission efficiency~\cite{20223DUAV}. The UAVs need to adjust their paths to avoid inter-UAV collisions and adapt to environmental uncertainties like obstacles and channel dynamics. The UAVs' trajectory planning is also related to the GUs' data transmission requirements, balancing the trade-off among coverage, energy efficiency, and communication performance. The UAVs' collaboration in data transmissions can be facilitated through the UAV-to-UAV (U2U) connections and transmission control. For example, the UAVs far away from the BS often suffer from poor channel conditions by using only the direct UAV-to-BS (U2B) channels, reducing the transmission efficiency and network capacity. Thus, it can be more efficient to exploit U2U links to relay data transmissions from the distant UAVs to the BS, forming a multi-hop relay network to support energy-efficient data sensing over a large-scale service area~\cite{2023Bayesianw}. Additionally, the UAVs' simultaneous data transmissions can lead to severe signal interference. The transmission control is further complicated by the UAVs' mobility and channel dynamics. Potentially, a higher network throughput can be achieved by dynamically optimizing the UAVs' collaborative transmission control and U2U relaying strategies along with the UAVs' trajectory planning strategy.

The spatio-temporal coupling among the UAVs' transmission control, U2U networking, and trajectory planning poses significant challenges for network performance maximization. It firstly relies on the availability of global network information and secondly calls for an efficient control algorithm, especially in large-scale UAV-assisted wireless networks. The absence of information exchange may lead to overlapped service areas among the UAVs, decreasing the network capacity and resource utilization~\cite{2023Bayesianw}. However, real-time information exchange among all UAVs is normally impractical or unaffordable. For example, the U2U and U2B connections can become unavailable due to severely attenuated channel conditions as two UAVs move far away along their trajectories. Such intermittent U2U and U2B connections thus introduce random information delay or loss~\cite{2024dt}. Additionally, the UAVs are required to simultaneously optimize the trajectories, transmission control, and resource allocation strategies, which is computationally demanding especially with dynamic channel conditions and information uncertainties. These motivate us to design robust and adaptive control algorithms, capable of handling incomplete network information while ensuring reliable and efficient network performance.

\vspace{-0.2cm}
\subsection{New Designs and Contributions}
In this paper, we first improve the UAVs' information sharing to enable more efficient collaboration. Typically, the high dimensional multiple UAVs' joint control can be formulated as a Markov decision process (MDP). It can be solved by a multi-agent deep reinforcement learning (MADRL) framework, which relies on full information sharing among all UAV agents. In particular, for each UAV, the decision-making agent should be aware of the complete and real-time network state and the other UAVs' actions during the centralized training phase~\cite{bai2023towards}. This implies that all UAVs should be fully connected to enable real-time information exchange. On the other hand, the UAVs' real-time information exchange consumes significant channel resources, sacrificing the network capacity and resource efficiency.

Therefore, it becomes a critical design problem to balance the UAVs' information exchange for efficient learning and data communications. We address this trade-off by designing a delay-tolerant MADRL framework for the UAVs' decision making in communication-limited scenarios. Although efficient information exchange via U2U connections is not always feasible, each UAV can update and cache its status information, i.e., service location, traffic demands, and channel conditions, at the BS when it reports the sensing data to the BS via the U2B connection. Meanwhile, each UAV can retrieve the other UAVs' status information from the BS's ACK packets without direct U2U connections~\cite{2024dt}. To reduce the information delay at the BS, we integrate a delay-penalized reward design into the MADRL framework, which encourages all UAVs to optimize their trajectories and maintain frequent information exchange with the BS, fostering awareness of the network state and more efficient UAVs' collaboration.

As the UAV-assisted wireless network scales up, each UAV increasingly relies on multi-hop U2U links to forward its data to the BS, which leads to larger information delays for decision-making. Such delayed information will misguide UAVs' trajectory planning and transmission control. As such, we aim to mitigate the impact of delayed information resulting from limited communication by leveraging historical information cached at the BS. Specifically, we integrate a spatio-temporal attention module into the MADRL framework so that each UAV can predict other UAVs' delayed information, and then adapt its trajectory and transmission control strategies accordingly. The prediction module exploits both the temporal correlations in the UAV's historical information and the spatial dependencies among neighboring UAVs. By estimating the delayed information, each UAV can better understand the complete network state and support more efficient collaboration without real-time inter-UAV information exchange.

We envision that the spatio-temporal attention enhanced MADRL (STA-MADRL) together with the delay-penalized reward design can provide a practical multi-agent decision-making framework for decentralized wireless networks. Different from the existing studies, we focus on communication-limited scenarios and design an incentive mechanism to promote timely information updates. Moreover, leveraging historical interactions for prediction enables effective collaboration even when the channels for information sharing are sporadic and error-prone. Specifically, our main contributions are summarized as follows:
\begin{itemize}
\item {UAVs' joint control in communication-limited scenarios:} Multiple UAVs are used to collect data from the GUs and forward the data to the remote BS. The UAVs' trajectory planning is firstly required to enhance network coverage. The UAVs further adapt the multi-hop network formation along the UAVs' trajectories. Besides, the UAVs optimize transmission control to better serve the GUs simultaneously. Different from the current literature, we focus on network throughput maximization in a communication-limited scenario, in which the UAVs' information sharing is unreliable and intermittent.
\item {Delay-penalized reward improving UAVs' information sharing:} Information sharing is required to guide the UAVs' efficient collaboration, while it also poses extra constraints on the UAVs' trajectory planning and limits the transmission capability in the communication-limited scenario. As such, we devise a delay-penalized reward in the MADRL framework that guides the UAVs' trajectory planning to maintain frequent information exchange without a significant loss in transmission capability.
\item {Spatio-temporal attention predicting UAVs' information loss:} The information delay or loss becomes inevitable when the network scales up. Instead of relying on frequent information exchange, we further propose the STA-MADRL algorithm to exploit the UAVs' historical observations by using a spatio-temporal attention based prediction approach, which enhances the UAVs' awareness of the complete network state and facilitates more efficient collaboration in the joint control. The simulation results verify that STA-MADRL achieves over 50\% reduction in information delay and 75\% throughput gain comparing to the communication-limited MADRL.
\end{itemize}

Some preliminary results of this work have been presented in a conference paper~\cite{2024dt}, which verifies the effectiveness of the delay-penalized reward to guide the UAVs' trajectory planning. In this paper, we focus on compensating the UAVs' excessive information delay as the UAV-assisted wireless network scales up, and integrate a spatio-temporal attention module into the MADRL framework to optimize the UAVs' trajectories, network formation, and transmission control strategies efficiently. The remainder of this paper is organized as follows. The literature review is provided in Section~\ref{sec-related}. We detail the system model in Section~\ref{sec-model} and propose the delay-tolerant MADRL for throughput maximization in Section~\ref{sec-dt}. Then, considering excessive information delay, we further propose the STA-MADRL framework in Section~\ref{sec-sta}. Finally, we present extensive results in Section~\ref{sec-sim} and draw the conclusions in Section~\ref{sec-con}.

\section{Related Works}\label{sec-related}
\subsection{Joint Trajectory Planning and Transmission Control}

Efficient multi-UAV deployment enables concurrent service for multiple GUs, improving network coverage, capacity, and transmission efficiency. The UAVs' fast and dynamic deployment also comes with the price for agile network adaptability that demands adaptive transmission control and networking strategies along with the UAVs' trajectories. The authors in~\cite{2021UAVBS} explored the UAVs' deployment and GUs' scheduling strategies to maximize network throughput while maintaining fairness. The authors in~\cite{2024WPCNKim} maximized the GUs' minimum uplink throughput by jointly optimizing the UAVs' trajectories and resource allocation strategies under energy neutrality and mobility constraints in wireless-powered networks. The authors in~\cite{2023MALQin} jointly optimized the UAV's trajectory planning, task offloading, and computing resource assignment to minimize system energy consumption of a UAV-based air-ground integrated computing network under limited battery capacity. UAVs' multi-hop relay network has also been considered in~\cite{chen2020joint} to enhance data transmission, task offloading, and on-the-fly computation. The authors in~\cite{2023Bayesianw} exploited the UAVs' energy-efficient network formation strategy to dynamically adapt multi-hop relay connections along with their trajectories. The joint optimization of the UAVs' trajectories and transmission control has also been proved indispensable for enhancing network performance in terms of fairness, energy efficiency, transmission delay, and energy consumption, e.g.~\cite{zeng2019energy,chen2020joint,qin2023multi}.

\subsection{Optimization vs. Learning for UAV-assisted Networks}
Network performance maximization in UAV-assisted wireless networks is typically solved either by optimization or machine learning methods. The authors in~\cite{2024HybridNOMA} employed the conventional block coordinate descent (BCD) method to derive a closed-form solution for transmit power that maximizes the minimum throughput in a UAV-assisted hybrid NOMA system. The authors in~\cite{2024JointSensing} proposed a generalized Dinkelbach and successive convex approximation algorithm to jointly optimize the transmit beamforming, computation offloading, and the UAV's trajectory in a joint sensing, communication, and computation framework. However, conventional optimization methods generally rely on real-time and global network information to perform heuristic decomposition, problem-specific simplification, and iterative approximation, leading to high computational complexity and even non-guaranteed performance. DRL can adapt to the changing network environment and learn optimal policies through continuously interactions with the network environment, making it particularly effective for trajectory planning, network formation, and transmission control in UAV-assisted wireless networks. The authors in~\cite{2022IOTCollisionAvoidance} proposed the dueling double deep Q-network (D3QN) algorithm for trajectory planning to maximize the collected data from multiple GUs under realistic constraints. The authors in~\cite{2024UGV} proposed the deep deterministic policy gradient (DDPG) algorithm for trajectory planning of both UAVs and unmanned ground vehicles (UGVs), which can supply energy to the UAVs. The authors in~\cite{2022vtTimelyData} employed a multi-agent DDPG (MADDPG) algorithm to jointly optimize the UAVs' trajectories and the expected age of information (AoI). Considering the UAVs' limited computation and energy capacities, the authors in~\cite{2022TaskOffloading} proposed the multi-agent twin-delayed DDPG (MATD3) algorithm to minimize the sum of execution delays and energy consumption in a task offloading system between multiple UAVs and edge nodes. A multi-agent advantage actor-critic (MAA2C) algorithm was used in~\cite{2022UAVDep} to optimize the UAVs' trajectories and enhance GUs' spectral efficiency for downlink transmissions. Though MADRL has been proven particularly effective in handling complex and dynamic systems, it still faces significant challenges for practical and flexible deployment in UAV-assisted wireless networks. The requirement of complete network information for centralized training becomes difficult when multiple UAVs have limited information about each other, leading to instability and slow convergence in the training phase. The intermittent channels among UAVs also make it costly for real-time information sharing. These challenges call for more efficient MADRL approaches for large-scale UAV-assisted wireless networks with limited communications.

\subsection{UAVs' Collaboration with Limited Communications} %

With limited information sharing, the authors in~\cite{jiang2018learning} proposed the attentional communication model (ACM) to selectively share critical information among agents, thus reducing the communication overhead. The authors in~\cite{li2021graph} proposed a graph neural network (GNN) based framework to compress and share only essential graph-structured information among UAVs. Federated learning can also be employed to enable the collaborative model training without raw data sharing~\cite{zhu2020federated}. The authors in~\cite{25message} formulated a multi-agent partially observable MDP to minimize the total energy consumption of a UAV-assisted MEC network by optimizing the UAVs' mobility, user association, resource allocation, and task offloading decisions. A neural network based interaction mechanism is integrated into the MADRL framework to help UAVs generate autonomously task-oriented messages for energy minimization. Our previous work in~\cite{2023Bayesianw} proposed a data-driven Bayesian optimization approach to guide MADDPG by estimating each UAV's best action in every decision epoch. Such a data-driven action estimation only focuses on the reward in the next step, and thus it is short-sighted and may be contradictory with the MADDPG's action.

Instead of relying on information sharing, recent research has also explored prediction-based approaches to estimate the lost or delayed information for UAVs' collaboration. The authors in~\cite{Wang2023} proposed a deep recurrent Q-network (DRQN) framework to optimize the UAV's trajectory by predicting the partially observable system state, enabling spectral-efficient resource allocation and scheduling. The authors in~\cite{Wu2024} proposed a GNN framework for the UAVs to predict the other UAVs' trajectories using historical information, suppressing the need for real-time information sharing. The authors in~\cite{24graph-uav} proposed a graph-attention multi-agent trust region reinforcement learning framework for trajectory planning and resource assignment in a multi-UAV-assisted wireless network. The graph recurrent network is used to process the network topology and extract useful information and patterns from historical observations. Different from existing works, in this paper we exploit the spatio-temporal dependencies from both the UAVs' trajectories and networking strategies to enhance UAVs' collaboration.

\section{System Model}\label{sec-model}

\begin{figure}[t]
	\centering
	\includegraphics[width=0.9\linewidth]{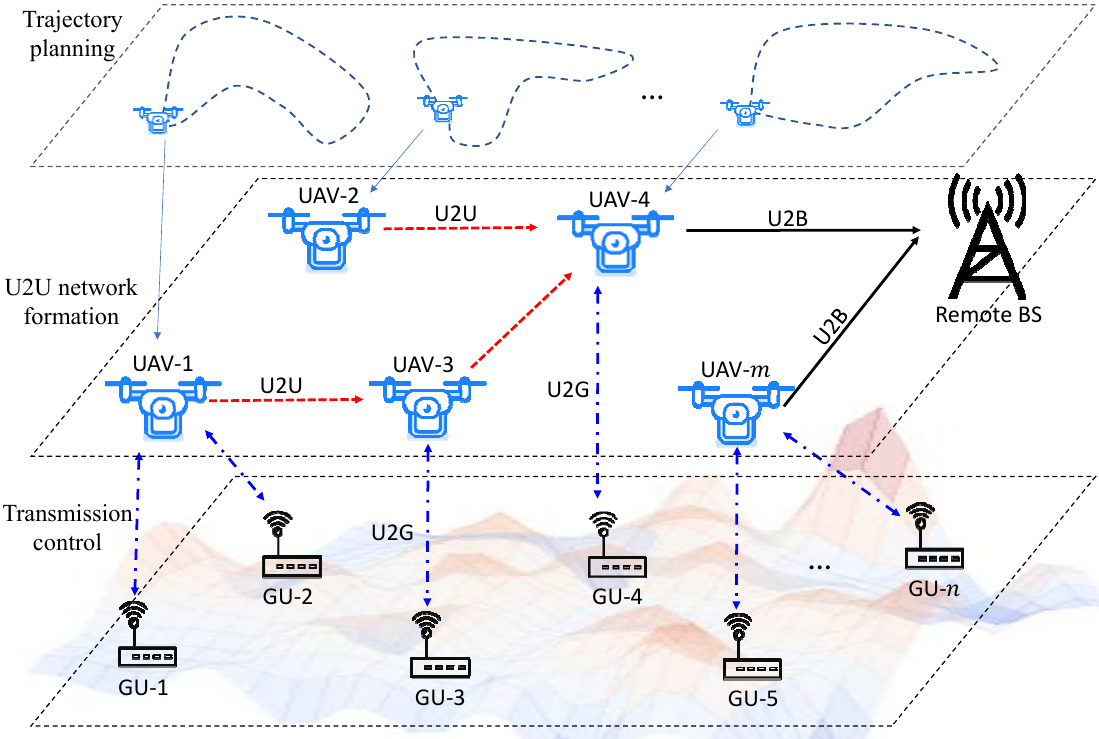}
	\caption{Multiple UAVs assist GUs' data tranmissions to the remote BS, by joint trajectory planning, network formation, and transmisison control.}
	\label{fig:system_model}
	\vspace{-0.2cm}
\end{figure}

Considering a large set of GUs randomly distributed over the service area, we employ multiple UAVs to assist GUs' data transmissions to the remote BS by optimizing the UAVs' trajectory planning, network formation, and transmission control strategies. As illustrated in Fig.~\ref{fig:system_model}, the multi-UAV-assisted wireless network consists of one remote BS, $N$ UAVs, and $M$ GUs, similar to those in~\cite{2023Bayesianw} and~\cite{2024dt}. The sets of UAVs and GUs are represented as $\mathcal{N} = \left\{ 1,2,\ldots,N \right\}$ and $\mathcal{M} = \left\{ 1,2,\ldots,M \right\}$, respectively. We assume that the GUs cannot be served directly by the BS due to the long distance or obstacles. The UAVs are equipped with $F$ antennas while the GUs are single-antenna devices. The UAVs' signal beamforming can be used to enhance the wireless data transmissions and the wireless energy transfer to sustain the low-power GUs. The whole system is operated in a time-slotted framework. Each time frame is divided into multiple time slots with a unit length. The set of time slots is denoted by $\mathcal{T}\triangleq\{1,2,\ldots, T\}$. The UAVs' operations in each time slot can be further divided into three phases, i.e.,~wireless energy transfer $t_{e}$, uplink data collection $t_{s}$, and data forward transmission $t_{r}$. Each UAV firstly moves to a fixed location and beamforms energy to the GUs in the first phase $t_{e}$, and then collects the sensing data from the GUs for the second phase $t_{s}$. After that, the UAVs forward the sensing data directly to the BS or relay it to the other UAVs via the U2U links in the third phase $t_{r}$.

\subsection{UAV-assisted Wireless Channel and Energy Transfer}

Let $\ell_0= (x_0, y_0, h_0)$ denote the BS's location with the fixed height $h_0$, which is viewed as the UAV-$0$. The set of all UAVs including the BS is defined as $\tilde{\mathcal{N}} \triangleq \mathcal{N} \cup\{0\}$.
The UAV-$n$'s trajectory  is defined as a collection of location points in different time slots, i.e.,~$\mathcal{L}_n=\{\ell_n(1), \dots, \ell_n(t), \dots,\ell_n(T)\}$, where $\ell_n(t)\triangleq (x_n(t), y_n(t), H)$ denotes the location with a {fixed} altitude $H$ in the $t$-th time slot. {To ensure safety}, all UAVs need to satisfy the distance and speed constraints:
\begin{subequations}\label{equ-cons-v}
\begin{align}
\setlength\abovedisplayskip{0.01pt}
&|| \boldsymbol{\ell}_n(t+1) - \boldsymbol{\ell}_n(t) || \leq v_{\max} {t_{e}}, ~\forall n\in \mathcal{N},\label{prob-uav-con-v}\\
&|| \boldsymbol{\ell}_n(t) - \boldsymbol{\ell}_{n{'}}(t) || \geq d_{\min}, ~\forall n,n{'} \! \in \! \mathcal{N},~ n  \neq  n{'}, \label{prob-uav-con-collision}
\end{align}
\end{subequations}
where $v_{\max}$ and $d_{\min}$ denote the maximum flying speed and the minimum safety distance between {two UAVs}, respectively.

The GU-$m$'s location is indicated by $q_m= (x_m, y_m, 0)$. Let $\mathbf{h}_{n, m}(t) \in \mathbb{C}^{F \times 1}$ denote the channel between the UAV-$n$ and the GU-$m$ at the $t$-th slot. The UAV-to-GU (U2G) channel can be modeled as follows:
\begin{equation}\label{equ-h}
\mathbf{h}_{n, m}(t)= \frac{\omega_0^{1/2}}{\left\|\ell_n(t)-q_m\right\|}   \tilde{\mathbf{h}}_{n, m}(t),
\end{equation}
where $\omega_0$ is the channel gain at unit distance and the small-scale fading channel $\tilde{\mathbf{h}}_{n, m}(t)$ is a combination of the line-of-sight (LoS) and the non-LoS components. It is clear that the U2G channel $\mathbf{h}_{n, m}(t)$ depends on the UAVs' trajectory planning. Each GU has a battery with finite capacity $E_{\max}$ and can harvest RF energy from UAVs' signal beamforming during the power transfer period $t_e$. Let $E_m(t)$ denote the GU-$m$'s battery level at the beginning of the $t$-th time slot. Let $\mathbf{w}_{n}^{e}(t) \in \mathbb{C}^{1 \times F}$ denote the UAV-$n$'s beamforming vector in the power transfer phase $t_e$. Hence, the energy harvested by GU-$m$ can be evaluated as~\cite{2021EEYang}:
\begin{equation}
e^{h}_m(t) = \sum_{n\in\mathcal{N}}  \eta_{e} \int_{0}^{{t_{e}}} |\mathbf{h}_{n, m}(t)\mathbf{w}_{n}^e(t)|^{2} \text{d} t ,
\end{equation}
where $\eta_{e}$ denotes the energy conversion efficiency and $\mathbf{w}_{n}^e(t)$ is limited by the power constraint $||\mathbf{w}_{n}^{e}(t) ||^{2} \leq p^e_{n}$. The GU-$m$'s energy consumption $e_m^c(t)$ for uplink data transmission via the U2G channel is assumed to be a linear function of the transmission time, i.e.,~$e_m^c(t) = p_{m}^s t_{s}$, where the transmit power $p_m^s$ depends on the GU-$m$'s transmission rate and the U2G channel condition. To ensure sustainable operations, the GU-$m$'s energy consumption is constrained as follows:
\begin{equation}\label{energy_constr}
 e_m^c(t)\leq \min \{E_m(t) + e^{h}_m(t), E_{\text{max}} \}.
\end{equation}
Thus, the GU-$m$'s battery level evolves as follows:
\begin{equation}
\begin{aligned}
E_m(t+1) = \min \left\{ E_{\text{max}}, E_{m}(t) + e_m^{h}(t) - e_m^c(t)\right\}.
\end{aligned}
\end{equation}

\subsection{GUs' Scheduling and Uplink Data Transmissions}
Each GU independently generates sensing data with the random size $q_m(t)$ per time slot, which remains in its data buffer until the GU is scheduled by UAVs for uplink data transmission during the second phase $t_{s}$. Let the binary variables $\bm \Psi = \{\psi_{m,n}(t)\}_{m \in \mathcal{M}, n \in \mathcal{N}, t \in \mathcal{T}}$ denote the GUs' association strategy, where $\psi_{m, n}(t)=1$ indicates the association between GU-$m$ and UAV-$n$ in the $t$-th time slot, and $\psi_{m, n}(t)=0$ otherwise. We assume that each GU can associate with at most one UAV per time slot~\cite{2023Bayesianw,2024dt}:
\begin{equation}\label{equ-G2U}
\sum_{n\in \mathcal{N}}\psi_{m, n}(t) \leq 1 ,  ~~ \forall m\in\mathcal{M} \text{ and } t \in \mathcal{T}.
\end{equation}
Let $\mathcal{\tilde{M}}(t) \triangleq \{m |~ \forall n, \psi_{m, n}(t)=1\}$ denote the set of all active GUs in the same time slot. These GUs may create interference to each other during their uplink data transmissions. Given the UAV-$n$'s receive beamforming vector $\mathbf{w}_{n}^{s}(t) \in \mathbb{C}^{1 \times F}$, the GU-$m$'s signal to interference plus noise ratio (SINR) at the UAV-$n$ is defined as:
\vspace{-0.1cm}
\begin{equation}\label{equ-gamma}
\gamma_{m,n} = \frac{p^s_{m}|\mathbf{h}_{m, n} \mathbf{w}_{n}^{s}|^{2} }{\sigma_n^2 + { \sum_{m' \in \mathcal{\tilde{M}} \setminus m }  p^{s}_{ m'} |\mathbf{h}_{m', n}\mathbf{w}_{ n}^{s}|^{2} } } ,
\end{equation}
where $\sigma_n^2$ denotes the noise power, and the second term in the denominator is the interference to the GU-$m$'s signal reception at the UAV-$n$. For simplicity, we omit the time index in~\eqref{equ-gamma}. Thus, the uplink data rate from the GU-$m$ to the UAV-$n$ is given by:
\vspace{-0.1cm}
\begin{equation}
d^s_{m, n}(t)= \psi_{m, n}(t) t_s \log \Big(1+\gamma_{m,n}(t) \Big).
\end{equation}
Let ${Q}_{m}(t)$ denote the GU-$m$'s data buffer size. Thus, the GU-$m$'s buffer dynamics can be described as follows:
\vspace{-0.1cm}
\begin{equation}\label{equ-gu-buffer}
{Q}_{m}(t+1) = \max\left\{ {Q}_{m}(t) - \sum_{n \in \mathcal{N}} d^s_{m, n}(t),0 \right\}  +  q_m(t).
\end{equation}

\subsection{Network Formation via U2U Relay Communications}
During the forward transmission phase $t_r$, UAVs can either transmit data directly to the BS or forward it to the next UAVs via the U2U links. When the direct U2B channel is unavailable, e.g.,~the UAV is far away from the BS, the UAV will establish the U2U connection to a nearby UAV with more preferable channel condition, instead of carrying the sensing data and flying closer to the BS for direct transmissions. The adaptive formation and evolution of U2U connections along the UAVs' trajectories can be viewed as the UAVs' network formation. In particular, let binary matrix $\bm \Phi= \{\phi_{n, n{'}}(t) \}_{n, n{'} \in \tilde{\mathcal{N}}, t \in \mathcal{T}}$ denote the UAVs' {network formation} strategy, similar to that in~\cite{2023Bayesianw}. We have $\phi_{n, n{'}}(t)=1$ if the {UAV-$n$} transmits data to the {UAV-$n{'}$} via the U2U channel, and $\phi_{n, n{'}}(t)=0$ {otherwise}. To mitigate interference, we assume that each UAV can forward data in at most one U2U channel each time:
\begin{equation}\label{equ-NF-limit}
\sum_{n{'} \in \tilde{\mathcal{N}}, n' \neq n} \phi_{n, n{'}}(t) \leq 1, ~~\forall n \in { {\mathcal{N}}} \text{ and } t \in \mathcal{T}.
\end{equation}
Let $p^e_{n}$ denote the UAV-$n$'s transmit power and $\mathbf{w}_{n}^{r}(t)$ be the normalized transmit beamforming vector in the forward transmission phase. Then, {the {SINR} at the receiving UAV-$n'$ from the transmitting UAV-$n$ is given by: }
\begin{equation}
\gamma_{n,n'}^{r}  = \frac{p^e_{n} |{\mathbf{h}_{n,n'} } \mathbf{w}_{n}^{r} |^{2} }{\sigma_{n'}^2 + \sum_{k \in \mathcal{N} \setminus n}  \phi_{k,n'}  p^e_{k} |\mathbf{h}_{k, n'} \mathbf{w}_{k}^{r} |^{2}},
\end{equation}
where the second term in the denominator denotes the interference from the other UAVs forwarding data simultaneously. We assume that the U2U channel $\mathbf{h}_{n,n'}$ follows a similar model to that of the U2G channel. Thus, the size of data forwarded from the {UAV-$n$} to the {UAV-$n{'}$} is given by:
\begin{equation}
d_{n, n{'}}^r(t)= \phi_{n, n{'}}(t) t_{r}\log \Big(1+ \gamma_{n,n{'}}(t) \Big),
\end{equation}
and the total data received by the UAV-$n$ is defined as:
\vspace{-0.1cm}
\begin{equation}
d^{c}_n(t) = \sum_{m \in \mathcal{M}} d^s_{m,n}(t)+ \sum_{n' \in \mathcal{N} \setminus n }d^r_{n',n}(t),
\end{equation}
which includes the data collected from the GUs directly and that forwarded by the other UAVs. Via the U2U channels, the UAV-$n$ can also forward a part of its data to the other UAVs, denoted as $d^o_n(t) = \sum_{ n{'} \in \mathcal{\tilde{N}} \setminus n } d^r_{n, n{'}}(t)$. Similar to the GUs' buffer dynamics in~\eqref{equ-gu-buffer}, the UAV-$n$'s data buffer status $D_n(t)$ in each time slot dynamically evolves as follows:
\begin{equation}\label{equ-data-upgrade}
D_n(t+1) = \max\left\{ D_n(t) - d^o_n(t), 0\right\} + d^c_n(t).
\end{equation}
Note that both GUs and UAVs may have limited data buffer size. Considering the low cost of buffer space, the buffer size can be practically very large comparing to the size of sensing data, and thus we omit the buffer limits in~\eqref{equ-gu-buffer} and~\eqref{equ-data-upgrade}.

\section{Delay-Tolerant MADRL for UAVs' Collaborative Control}\label{sec-dt}

We aim to maximize the network capacity by jointly optimizing the UAVs' trajectory planning $\mathcal{L} \triangleq \{\mathcal{L}_n\}_{n\in\mathcal{N}}$, network formation $\bm \Phi$, and transmission control strategies $(\bm W,\bm \Psi)$, including the UAVs' beamforming $ \bm W \triangleq \left\{ \mathbf{w}_n^{e}(t),\mathbf{w}_n^{s}(t), \mathbf{w}_n^{r}(t) \right\}_{n \in \mathcal{N}, t\in \mathcal{T}}$ and the GUs' scheduling strategy $\bm \Psi$, which can be formulated as:
	\begin{equation}\label{prob-energy-efficiency}
        \max_{\mathcal{L}, \bm \Phi, \bm W, \bm \Psi} ~ \mathbb{E}\left[\sum_{n\in\mathcal{N}} d^r_{n,0}\right] \quad  \text{s.t.} \,\, ~\eqref{equ-cons-v}-\eqref{equ-data-upgrade}.
	\end{equation}
The expectation is over different time slots $t\in\mathcal{T}$ and $d^r_{n,0}$ denotes the data forwarded to the remote BS by the UAV-$n$ per time slot. Problem~\eqref{prob-energy-efficiency} is a mixed-integer nonlinear program and challenging to solve. The UAVs' network formation $\bm \Phi$ and GUs' scheduling $\bm \Psi$ are discrete variables, both scaling exponentially with the size of the GUs and UAVs. Additionally, different UAVs' trajectory planning and beamforming strategies are strongly coupled in time and space domains, which makes one-shot optimization computationally extensive. In the sequel, we first propose the general MADRL framework to solve problem~\eqref{prob-energy-efficiency} with complete network information. Then, we reveal the practical limitations of MADRL in large-scale UAV-assisted wireless networks due to the unreliable channels for UAVs' information exchanges. This motivates us to propose the delay-tolerant MADRL algorithm to guide the UAVs' trajectory planning that improves both information sharing and network capacity.

\subsection{General Multi-agent DRL Framework}

The effective application of DRL approaches depends on a proper reformulation of problem~\eqref{prob-energy-efficiency} into MDP, which offers a robust mathematical modeling for sequential decision-making problems. For a multi-UAV-assisted wireless network, the system state at each time slot consists of all UAVs' observations $\bm o(t) \triangleq \big\{ \bm\ell (t), \bm D (t) \big\}$, including all UAVs' locations $\bm\ell (t) \triangleq \{\bm \ell_n(t)\}_{n \in \mathcal{N}}$ and their buffered data $\bm D(t) \triangleq \{D_n (t)\}_{n \in \mathcal{N}}$. According to the optimization problem in~\eqref{prob-energy-efficiency}, the UAV-$n$'s action in the $t$-th time slot $\bm a_n(t)$ includes {the next trajectory point ${\ell_n(t+1)}$}, the U2U network formation $\{\phi_{n,n'}(t)\}_{n'\in\tilde{\mathcal{N}}}$, the beamforming vectors $\{ \mathbf{w}_n^{e}(t),\mathbf{w}_n^{s}(t), \mathbf{w}_n^{r}(t) \}$, and the GUs' association strategy $\{\psi_{m,n}(t)\}_{m\in\mathcal{M}}$. The UAVs' joint action is given by $\bm a(t) \triangleq \{\bm a_n(t)\}_{n \in \mathcal{N}}$. The reward depends on the complete network state and all UAVs' actions. Considering the throughput maximization in~\eqref{prob-energy-efficiency}, we assume that each UAV shares the same reward function $R(t)$, defined as follows:
\vspace{-0.2cm}
\begin{equation}\label{equ-reward}
R(t)=  \sum_{n=1}^{N}  \mu_1 d^r_{n,0}(t) + \mu_2 Z_{n}(t) ,
\end{equation}
where $\mu_1$ and $\mu_2$ are the weighting parameters for the overall throughput and a penalty term $Z_{n}(t)$, respectively. The transmission reward $d^r_{n,0}(t)$ encourages each UAV-$n$ to forward more information to the BS, while the penalty term $Z_{n}(t)$ will ensure the UAVs' safety during trajectory planning:
\vspace{-0.1cm}
\[
Z_{n}(t) = \sum_{n,n^{'} \in \mathcal{N}} \Big(\mathbf{I}(\hat{\ell}_n(t) \leq v_{\max})+ \mathbf{I}\left(\ell_{n, n'}(t)\geq d_{\min }\right)\Big),
\]
where $\hat{\ell}_n(t)$ is the UAV-$n$'s speed and $\ell_{n, n'}(t)$ is the distance between two UAVs. Here $\mathbf{I}(\cdot)$ denotes the indicator function.

Each agent in the MADRL framework has distributed decision-making capabilities, dynamically adjusting its next action based on its perception of the network state using a pair of deep neural networks (DNNs), namely, the actor and the critic networks. The actor network selects actions by approximating the policy function, while the critic network evaluates these actions by estimating the value function. These two networks enable each agent to adapt and evaluate actions during the training phase until all agents' joint policies stabilize with the highest reward value. Let $\theta_t$ and $w_t$  denote the DNN parameters of the actor- and critic-network, respectively. Given the global network information $\bm{o}(t)$, the UAV-$n$'s actor-network $\Pi_n^{\theta_t}$ aims to generate the action $a_{n}(t) = \Pi_n ^{\theta_t}\left(\bm{o}(t)\right)$ to maximize the value function $\mathcal{J}_{n}\left( \theta_t  \right)$, which is defined as the cumulative discounted reward $\mathcal{J}_{n}\left( \theta_t  \right) =\mathbb{E}_{a}\left[\sum_{t=0}^{\infty} \rho_{r}^t R_{n}(t) \right]$, where $\rho_{r}$ represents the discount factor for the rewards in different time slots. However, the true value function $\mathcal{J}_{n}\left( \theta_t  \right)$ is unattainable during online learning due to limited sample size. Thus, the critic-network $Q_n^{w_t}$ is further employed to approximate the true value function, namely Q-value, which is used to evaluate the quality of the state-action pair $\left( \bm{o}(t) , a_{n}(t)\right)$. A large Q-value implies that the action $a_{n}(t)$ can be more preferable when visiting the same state in the future time steps. To this end, we can update the policy $\Pi_n ^{\theta_t}$ by using the deterministic {policy gradient method~\cite{2017maddpg_UAV}}. The critic-network $Q_n^{w_t}$ can be updated according to the temporal-difference (TD) error between the online critic-network $Q_n^{w_t}$ and its target $y_n(t) = R_n(t) + \rho_{r} \hat{Q}_n^{\hat{w}_t}$. Here, the DNN parameter $\hat{w}_t$ of the target critic-network $\hat{Q}_n^{\hat{w}_t}$ is a delayed copy from ${w}_t$ to prevent an excessive overestimation of the value function. To minimize the TD error, the gradient descent method can be used to update the critic-network $Q_n^{w_t}$. More detailed derivations of the general MADRL framework can be referred to~\cite{2023Bayesianw} and thus omitted here for conciseness.

\vspace{-0.2cm}
\subsection{Delay-Tolerant MADRL with Limited Communications} \label{DPM}

Each UAV-$n$'s actor network in the general MADRL framework relies on its local observation $\bm o_n(t)$ and the other UAVs' observations $\bm{o}_{-n}(t)$ as input to generate the action $\bm a_{n}(t) = \Pi_n ^{\theta_t}\left(\bm o_n(t),\bm{o}_{-n}(t)\right)$. This requires that all UAVs can report their real-time local information to the BS and share the complete network state $\bm{o}(t)=(\bm o_n(t),\bm{o}_{-n}(t))$ with all UAVs for efficient coordination. We denote it as the Ideal-MADRL algorithm in the following discussions. The Ideal-MADRL's real-time environmental awareness allows different UAVs to plan their optimal trajectories and transmission control decisions to avoid signal interferences and inefficiency.

However, when the UAVs' real-time information exchange becomes unavailable in a practical system, the learning performance may degrade significantly due to the misaligned information input $(\bm o_n(t),\tilde{\bm{o}}_{-n}(t))$ to the UAV-$n$'s actor network, where $\tilde{\bm{o}}_{-n}(t)$ can be viewed as the delayed network information from the other UAVs. For example, some UAVs may be distant from the BS and thus the U2B links are not available. As such, they cannot send their local information to the BS via U2B links in every time slot. Two UAVs may also be disconnected as they fly away from each other, making it difficult for their information exchange via the U2U link.

\subsubsection{Delay-penalized reward design} Considering the practical challenges for real-time information exchange in UAV-assisted wireless networks, we are motivated to design a delay-tolerant MADRL algorithm that accounts for the discrepancy between $\tilde{\bm{o}}_{-n}(t)$ and ${\bm{o}}_{-n}(t)$. To achieve this, we record the actual time delay between $\tilde{\bm{o}}_{-n}(t)$ and ${\bm{o}}_{-n}(t)$ in each time slot, and propose a delay-penalized reward function for each UAV to guide its trajectory planning. In particular, let $\zeta_n(t)$ denote the UAV-$n$'s information delay since its last U2B connection with the BS. When the time delay $\zeta_n(t)$ becomes large, the delayed information $\tilde{\bm{o}}_{-n}(t)$ will be very different from the real state ${\bm{o}}_{-n}(t)$. Hence, we employ the time delay $\zeta_n(t)$ as a penalty term that forces the UAV to fly closer to the BS and report its newest network information via the U2B link, avoiding a continuous increase in the information delay. As such, we can simply revise the reward in~\eqref{equ-reward} as follows to account for the UAVs' information delay:
\vspace{-0.1cm}
\begin{equation}\label{equ-reward-delay}
 \widetilde R(t)= \sum_{n=1}^{N}  \mu_1(\omega_1d^r_{n,0}(t) - \omega_2\zeta_n(t))+ \mu_2 Z_{n}(t),
\end{equation}
where $\omega _1$ and $\omega_2$ are the non-negative weighting coefficients that balance network throughput and information delay, respectively. This new reward function encourages more frequent interactions between UAVs and the BS. When all UAVs' information delays are small, the complete network information can be timely shared among all UAVs, leading to a fully coordinated UAV-assisted wireless network to serve the GUs with minimum conflicts and interferences.

\subsubsection{Evaluating the information delay}
We assume that the BS can cache the latest state information reported by the UAVs via the U2B links. For each UAV-$n$, when there exists the U2B link, it can forward the sensing information to the BS and report its local state information $\bm o_n(t)$ as well, which will replace the obsolete information cached by the BS. The UAV-$n$ can also retrieve all state information ${\bm{o}}_{-n}(t)$ of the other UAVs via the U2B link. When the UAV-$n$ is far from the BS and there is no U2B link, the BS will maintain the same state information $\bm o_n(t)$ for the next time slots. The UAV-$n$ cannot retrieve the latest information from the BS and have to rely on delayed information to make decisions.

In the $t$-th time slot, let $c_{n}(t)$ denote the most recent time slot when the UAV-$n$ exchanges state information with the BS to report its local information and retrieve the global network information. Note that the information exchange depends on the availability of the U2B link. We can update $c_{n}(t)$ as:
\begin{align}\label{equ-cn}
c_{n}(t) = c_{n}(t-1) (1-\phi_{n, 0}(t)) + t \phi_{n, 0}(t).
\end{align}
When the U2B link exists, i.e.,~$\phi_{n, 0}(t) = 1$, the UAV-$n$ will report its information to the BS and thus we have $c_{n}(t) = t $ and otherwise $c_{n}(t) = c_{n}(t-1)$. Hence, the information delay $\zeta_n(t)$ can be simply evaluated as $\zeta_n(t) =  t - c_{n}(t)$, which represents the number of time slots since the UAV-$n$'s last successful U2B connection with the BS. In this case, the BS maintains the obsolete state information $\bm o_n(t-\zeta_n(t))$ to approximate the UAV-$n$'s most recent state information $\bm o_n(t)$ in the $t$-th time slot. When the UAV-$n$ reports its state information to the BS in the $t$-th time slot, the other UAVs can immediately access its latest state information $\bm o_n(t)$ with zero delay, i.e.,~$\zeta_n(t) = 0$. Otherwise, they may make biased decisions based on the obsolete state information $\bm o_n(t-\zeta_n(t))$. It is clear that the UAV-$n$'s information delay $\zeta_n(t)$ depends on the frequency of U2B connections with the BS. Infrequent connections force the UAVs to use delayed information for trajectory planning and network formation, potentially leading to interference or service conflictions.

\section{Spatio-Temporal Attention Enhanced MADRL}\label{sec-sta}

Delay-tolerant MADRL records all UAVs' information delay and relies on the delay-penalized reward to guide the UAVs' trajectory planning that ensures frequent information exchange with the BS. As such, all UAVs' information delay can be maintained at a low level. However, as the network scales up with more UAVs, it becomes more difficult to plan all UAVs' trajectories and schedule their information exchange with the BS while maintaining the low information delay. Note that the U2B connections are available only when the UAVs are close to the BS. The U2B transmissions can be congested and thus the information delay will inevitably increase as the number of UAVs increases. Besides the delay-penalized reward design, we further explore how UAVs can fly efficiently to counter the inevitable information delay or loss as the network scales up. Our basic idea is to predict and recover the UAVs' missing information, instead of acquiring it from the BS via U2B connections, by exploiting the spatio-temporal dependencies in the UAVs' historical information.

\begin{figure}[t]
	\centering
	\includegraphics[width=1\linewidth]{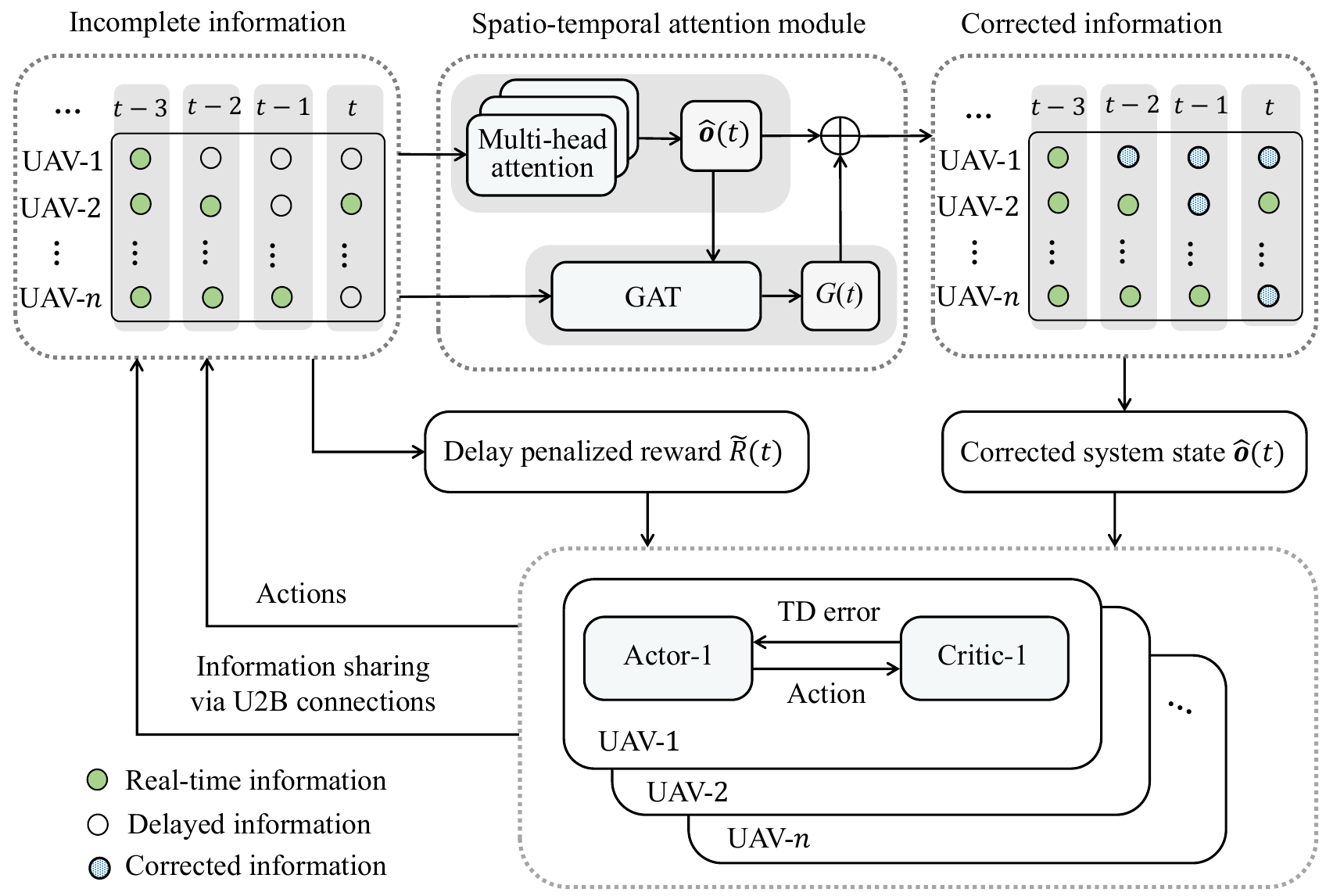}
	\caption{The STA-MADRL framework.}
	\label{fig-sta-frame}
	\vspace{-0.2cm}
\end{figure}

\vspace{-0.2cm}
\subsection{Spatio-Temporal Attention for Information Prediction}\label{title:ST}

The overview of the proposed STA-MADRL framework is shown in Fig.~\ref{fig-sta-frame}, which incorporates the delay-penalized reward design to improve the UAVs' information exchange with the BS, and leverages a spatio-temporal prediction module to recover the missing information and thus improve the UAVs' awareness of the complete system state for collaborative trajectory planning and transmission control.

The spatio-temporal prediction module combines temporal and spatial attention mechanisms to improve state prediction. The temporal attention module predicts the latest state information by analyzing individual UAV's historical behavior pattern. Temporal dynamics are captured using multi-head attention (MHA)~\cite{2017attention}. The spatial attention module exploits different UAVs' spatial correlations for information prediction using graph attention networks (GAT)~\cite{24graph-uav}. The combination of spatial and temporal features creates more comprehensive feature representations. The corrected state information is then input to the MADRL framework for each UAV to learn its trajectory planning, network formation, and transmission control strategies. As shown in Fig.~\ref{fig-sta-frame}, the BS maintains a table of the UAVs' state information, in which we use solid green points to denote the UAVs' real-time state information, while the hollow points represent the obsolete or missing information. Due to the UAVs' limited U2B communications with the BS, the state information cached by the BS is incomplete. Once some UAV has the U2B connection with the BS, its state information can be updated into the BS's information table. By counting each UAV's information delay, the delay-penalized reward can be evaluated to assess the quality of the UAVs' current actions, including the trajectory planning and transmission control strategies. The BS's information table can be further processed by the spatio-temporal attention module, which corrects the delayed information or fills up the missing entries in the information table. The corrected information table then provides the complete system state for all UAVs to make coordinated decisions in the MADRL framework. In the sequel, we present the details of the spatio-temporal attention based prediction module.

\vspace{-0.2cm}
\subsection{Temporal Multi-head Attention}\label{Title:Temporal}

The temporal attention aims to predict each UAV's future state information by exploiting the dependencies in its historical information. For example, when the GUs' traffic demands exhibit a stable spatial distribution over time, the UAVs may also have stable and periodically repeated trajectories to serve all GUs, which can be exploited for predicting the UAVs' future trajectories. To proceed, we define a moving window with length $\tau_0$ to predict the UAV's next state information by focusing on the most recent state information in the past $\tau_0$ time slots. This allows the temporal prediction model to capture the recent trends efficiently.

At the $t$-th time slot, we collect the UAV-$n$'s historical state information from time slot $t-\tau_0$ to $t-1$, denoted as $\mathcal{X}_n(t) = \{ \bm o_{n}(t-\tau_0), \dots, \bm o_{n}(t-1)\} \in \mathbb{R}^{\tau_{0} \times d_o} $, where $d_o$ represents the dimension of the UAV-$n$'s local state information, including its trajectory point and the buffer size. This sequence $\mathcal{X}_n(t)$ is then used to predict the current state $\hat{\bm o}_{n}(t)$. We implement the prediction by the MHA mechanism~\cite{2017attention}, which processes temporal dependencies in parallel, exhibiting prominent advantages over sequential algorithms such as long short-term memory (LSTM) and gated recurrent unit (GRU) in capturing nonlinear and long-range patterns. Specifically, MHA is composed of a set of DNNs that transform input sequences into queries $\mathbf{Q}$, keys  $\mathbf{K}$, and values $\mathbf{V}$, through linear projections using tunable weight matrices. It computes attention score using scaled dot-product independently in each head, allowing simultaneous focus on different feature subspaces. The output of each head is concatenated into a vector and then decoded as the final output of MHA. Such a multi-head architecture can capture complex temporal dependencies in high-dimensional feature space, making it particularly effective for processing sequential time series.

\begin{figure}[t]
	\centering
	\includegraphics[width=1\linewidth]{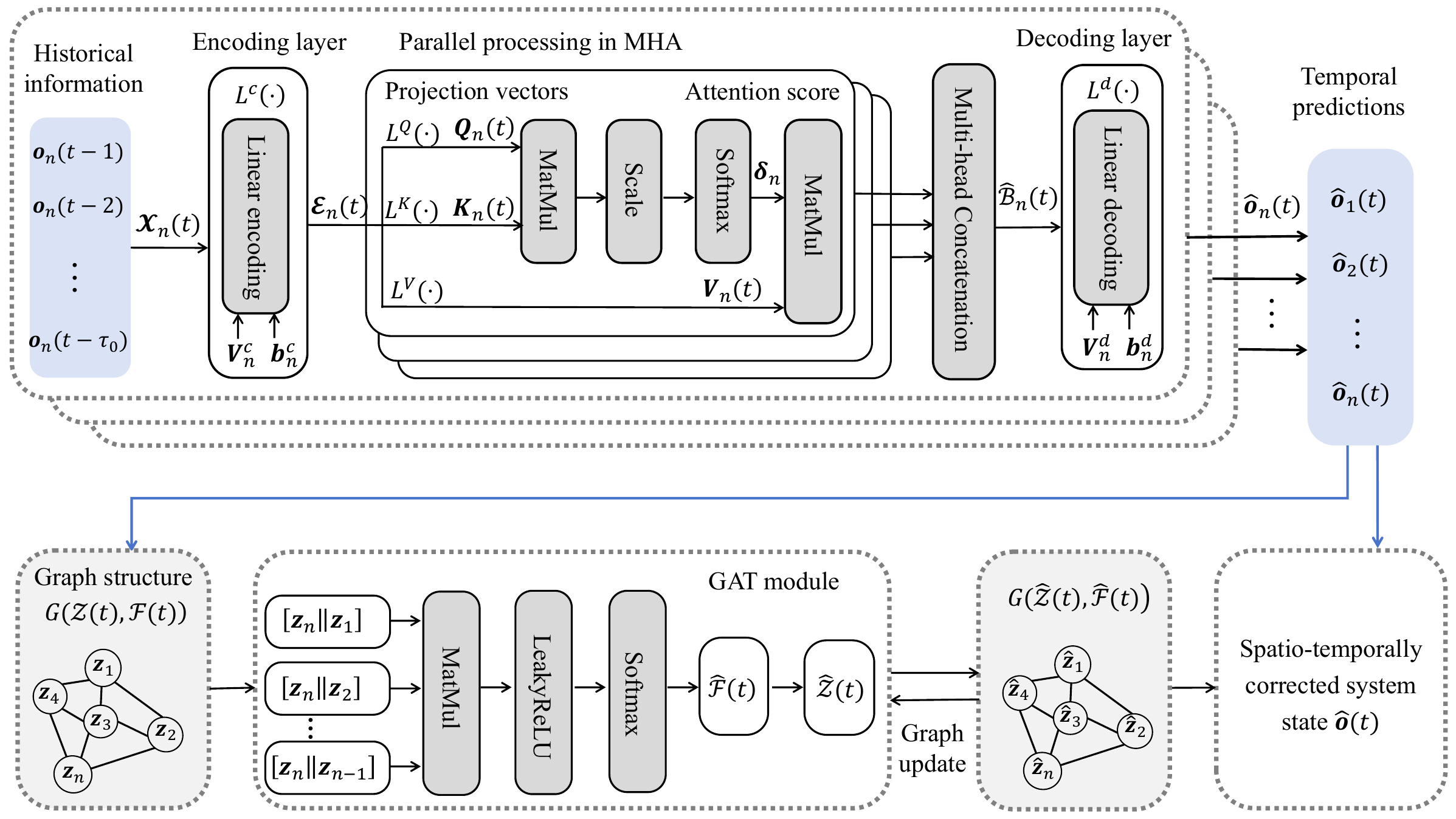}
	\caption{Network structure of the spatio-temporal attention prediction module.}
	\label{fig:method_model_illustration_module}
    \vspace{-0.2cm}
\end{figure}
As detailed in Fig.~\ref{fig:method_model_illustration_module}, the historical input sequence $\mathcal{X}_{n}(t)$ is firstly encoded by $L^{c}(\cdot)$ and linearly transformed into a high-dimensional form $\mathcal{E}_{n}(t)= L^{c}\left( \mathcal{X}_{n}\right)=  \mathcal{X}_{n}(t) (\bm V_n^c)^T +\boldsymbol b_n^c \in \mathbb{R}^{\tau_{0} \times d_{e}}$, where $\bm V_n^c \in \mathbb{R}^{d_{e} \times d_{o}} $ and $\boldsymbol b_n^c \in \mathbb{R}^{\tau_{0} \times d_{e}}$ are the weight matrix and bias of the encoding layer, respectively. For notational convenience, we omit the time index for the following discussions. After encoding, the dimension of the feature space increases from $d_o$ to $d_{e}$. The encoding into a high-dimensional space typically offers richer features for extraction and facilitates more efficient training of the subsequent attention modules. We then feed the encoded data $\mathcal{E}_{n}$ into parallel self-attention modules, using scaled dot-product computations to capture the correlations in the encoded data. Through three fully connected layers, the self-attention module maps $\mathcal{E}_{n}$ into query space $\mathbf{Q}_{n} =L^Q\left( \mathcal{E}_{n} \right) \in \mathbb{R}^{\tau_{0} \times d_{k}}$, key space $\mathbf{K}_{n} =L^K\left(\mathcal{E}_{n} \right) \in \mathbb{R}^{\tau_{0} \times d_{k}}$, and the value space $\mathbf{V}_{n} =L^V\left(\mathcal{E}_{n} \right) \in \mathbb{R}^{\tau_{0} \times d_{v}}$, respectively, where $d_{k}$ and $d_{v}$ represent the dimensions of the query-key space and the value space, respectively. The attention score of each query-key pair is $ \boldsymbol \lambda_n = \mathbf{Q}_{n}  \mathbf{K}_{n}^T $, which indicates query-key similarity, i.e.,~a higher value shows a stronger correlation. The scaled scores $\hat{\boldsymbol\lambda}_n = \frac{\boldsymbol \lambda_n }{\sqrt{d_{k}}}$ weight the value to focus on the most pertinent parts of the input for accurate predictions.

Subsequently, we apply the softmax function to transform the scaled scores $\hat{\boldsymbol\lambda}_n$ into attention matrix as follows:
\begin{align}\label{equ-attention-weight}
   \boldsymbol \delta_{n} =\operatorname{softmax}(\hat{\boldsymbol\lambda}_n ) ,
\end{align}
which represents the importance of the key $\mathbf{K}_{n}$ to the query $\mathbf{Q}_{n}$. A higher weight indicates that the key is more relevant. The final value is $\mathcal{B}_{n} = \boldsymbol \delta_{n}\mathbf{V}_{n} $. Furthermore, multiple independent attention operations in different heads are employed to stabilize the self-attention learning process. The outputs of all heads are concatenated as a vector $\hat{\mathcal{B}}_{n}$, which is then decoded as the predicted information vector as follows:
\begin{align}
\hat{{\mathcal{Y}}}_{n}(t) & = L^{d}(\hat{\mathcal{B}}_{n}) =  \hat{\mathcal{B}}_{n} (\bm V_n^d)^T +\boldsymbol{b}_n^d \nonumber\\
 & =  \{ \hat{\bm o}_{n}(t-\tau_0+1), \dots, \hat{\bm o}_{n}(t) \} \in \mathbb{R}^{\tau_{0} \times d_{o}},
\end{align}
where $\bm V_n^d \in \mathbb{R}^{d_{o} \times d_{v}} $ and $\boldsymbol b_n^d \in \mathbb{R}^{\tau_{0} \times d_{o}} $ are the weight matrix and bias of the decoding layer, respectively. Till this point, we can extract the UAV-$n$'s information $\hat{\bm o}_{n}(t)$ from $\hat{{\mathcal{Y}}}_{n}(t)$ at the $t$-th time slot. By collecting all UAVs' predictions $\{ \hat{\bm o}_{n}(t) \}_{n \in \mathcal{N}}$, the BS can build the complete state information.

\vspace{-0.2cm}

\subsection{{Graph-based Spatial Attention}}\label{Title:Spatial}

The temporal MHA focuses on the dynamic state transitions and correlations hidden in individual UAV's historical information, without considering cross-UAVs' spatial dependencies. In fact, two UAVs may have highly correlated observations when they are close to each other and have similar trajectories. Intuitively, it is feasible to predict one UAV's state based on the states of its neighboring UAVs. Such proximity-based spatial correlations imply that we may evaluate the spatial attention based on the distances among UAVs or the similarity of their trajectories. Considering the UAVs' mobility, we can model the UAVs' network formation in a dynamic graph structure, based on which we can perform graph computation and extract the spatial correlations among UAVs. We envision that the combination of temporal and spatial correlations can provide more accurate and comprehensive predictions for the UAVs' missing state information.

As shown in Fig.~\ref{fig:method_model_illustration_module}, we define an attention graph $G(\mathcal{Z}(t), \mathcal{F}(t))$ to represent the UAVs' spatial connections in the $t$-th time slot, where $\mathcal{Z}(t)   = \left\{ \bm z_{n}(t) \right\}_{n \in \mathcal{N}} $ denotes the set of nodes and $\mathcal{F}(t) = \left\{ f_{n,n^{\prime}}(t) \right\}_{n\neq n', n,n' \in \mathcal{N}} $ denotes the weights on edges connecting the nodes. Each node $z_{n}(t)\in \mathcal{Z}(t) $ corresponds to a UAV and contains the encoded observations of the UAV, i.e.,~$\bm z_{n}(t) = L^{s}( \hat{\bm  o}_{n}(t)) \in \mathbb{R}^{d_{s}} $. The encoding function $L^{s}( \cdot )$ maps the UAV's state information into the $d_{s}$-dimensional feature space, preserving the node's information for further graph computations. The edge weight $f_{n,n^{\prime}}(t)$ is the attention score between two UAVs evaluated based on their distance and U2U connections.

Given the graph structure $G(\mathcal{Z}(t),\mathcal{F}(t))$, we further extract spatial features by GAT module, which includes stacked graph attention layers. For each edge, the weight $f_{n,n^{\prime}} (t)$ can be updated by evaluating the similarity between two nodes:
\begin{equation}\label{equ-attention-weight-gat}
\hat f_{n,n^{\prime}}(t) =\operatorname{softmax}\Big(\operatorname{LeakyReLU}(\boldsymbol \rho [\bm z_{n}(t) \Vert\bm z_{n^{\prime}}(t)])\Big),
\end{equation}
where $[\bm z_{n}(t) \Vert\bm z_{n^{\prime}}(t)] $ denotes the concatenation of two vectors $\bm z_n(t)$ and $\bm z_{n^{\prime}}(t)$, and $\boldsymbol \rho $ is the GAT's trainable parameter. After updating all edges' weights in $\widehat{\mathcal{F}}(t)$, we can further update each node's feature by aggregating as follows:
\vspace{-0.2cm}
\begin{equation}
\hat{\bm z}_{n}(t) = \operatorname{softmax}\left( \sum_{n^{\prime} \in \mathcal{N} \setminus \{n\} } \hat f_{n,n'}(t) \bm z_{n^{\prime}}(t)\right).
\end{equation}
After iterative graph computation, the graph attention layer produces the final attention graph ${G}(\widehat{\mathcal{Z}}(t),\widehat{\mathcal{F}}(t))$.

The integration of the temporal and spatial attention module is shown in Fig.~\ref{fig:method_model_illustration_module}. Initially, the BS replaces the UAVs' obsolete state information $\bm o_{n}(t-\zeta_{n})$ with the predicted state information $\hat{{\bm o}}_n(t)$ by using the temporal attention module. The temporal predictions $\hat{{\bm o}}(t)$ then drive the spatial attention module to process and extract the dependency information from the graph ${G}(t)$. {The final attention graph ${G}(t)$ and the temporal prediction $\hat{\bm o}(t)$ are flattened into a one-dimensional vector by the fusion layer to generate the compensated and complete state information. Leveraging the enhanced state representation, MADRL can be employed to solve the throughput maximization problem, achieving preferable trajectories, network formation, and transmission control strategies. In summary, the proposed STA-MADRL framework for UAVs' collaboration addresses the practical challenges of the conventional MADRL framework from two aspects. Firstly, considering the UAVs' limited communications, we devise the delay-penalized reward to guide the UAVs' trajectory planning that ensures frequent information exchange with the BS. Secondly, we focus on the hidden spatio-temporal dependencies of the UAVs' historical observations and propose a prediction module to recover the complete state information for the UAVs' decision making in the MADRL framework.

\vspace{-0.2cm}

\section{Numerical Results}\label{sec-sim}

In this part, we numerically evaluate the STA-MADRL framework to demonstrate its performance in communication-limited UAV-assisted wireless networks. We consider $M=9$ GUs, $N=3$ UAVs, and one BS within a 2$\times$2 km$^2$ area. The $x$-$y$ coordinate is scaled to the range of $[-1, 1]$. The BS is located at $\ell_0=(1,1,0)$ while the GUs are randomly distributed. The UAVs can start services from arbitrary locations, operating at a fixed altitude of $H=100$ meters with a maximum speed of $v_{\text{max}}=20$ m/s. The main parameters are similar to those in~\cite{2023Bayesianw}.


\subsection{Information Sharing Improves Learning Performance}


We implement the delay-tolerant MADRL with the delay-penalized reward and compare it with Ideal-MADRL. Note that Ideal-MADRL assumes complete and real-time information available to all UAVs and thus it can serve as a theoretical performance benchmark. We also implement the conventional communication-limited MADRL, which relies on incomplete state information for decision making. Dictated by~\eqref{equ-reward-delay}, the delay-tolerant MADRL leverages both the throughput performance and delay statistics to guide the UAVs' trajectory planning that ensures frequent information exchange with the BS. As shown in Fig.~\ref{fig-conv-thu-delay}(a), the delay-tolerant MADRL algorithm demonstrates a higher convergence speed and improved reward performance compared with the communication-limited MADRL, verifying the importance of information sharing for the UAVs' efficient collaboration. By forcing the UAVs' frequent information sharing, we not only achieve a better learning efficiency, but also improve the overall throughput performance. However, compared with Ideal-MADRL, the throughput of the delay-tolerant MADRL still has a significant drop, implying that the importance of information sharing can be further exploited by more sophisticated algorithm design. In Fig.~\ref{fig-conv-thu-delay}(b), we show the average information delay in the delay-tolerant and communication-limited MADRL algorithms. It is obvious that the delay-tolerant MADRL achieves a lower information delay than that of the communication-limited MADRL. This verifies that the delay-penalized reward encourages the UAVs to have more frequent information exchanges with the BS. As such, the BS can timely update the UAVs' local state in the information table and share the latest system state with the other UAVs via ACK packets. With reduced information delay, all UAVs can make informative control decisions in the MADRL framework to achieve an improved throughput performance as revealed in Fig.~\ref{fig-conv-thu-delay}(a).

\begin{figure}[t]
	\centering
	\subfigure[Throughput dynamics.]{
		\begin{minipage}[b]{0.46\linewidth}
			\includegraphics[width=1\textwidth]{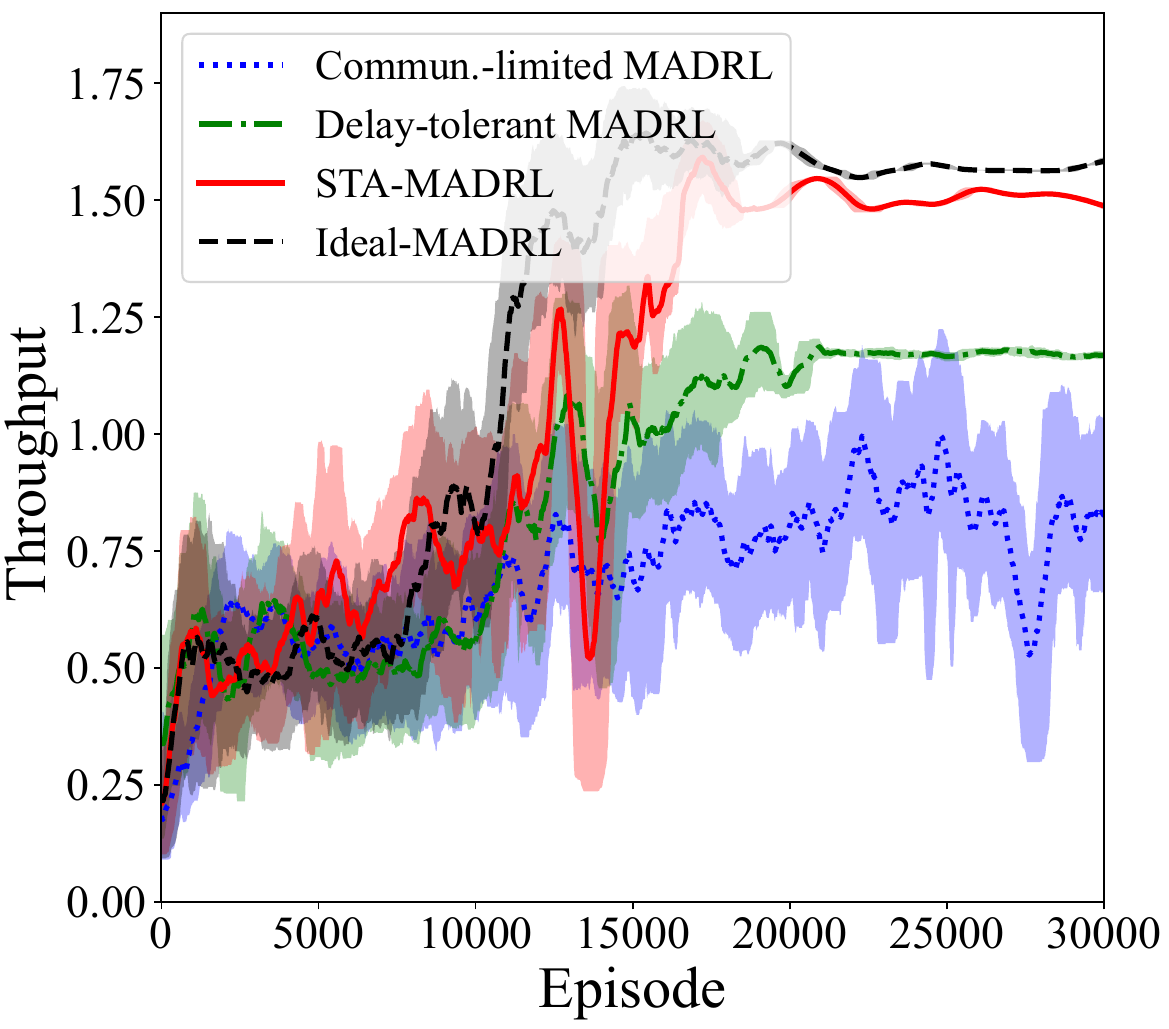}
		\end{minipage}
	}
	\subfigure[Dynamics of information delay.]{
		\begin{minipage}[b]{0.46\linewidth}
			\includegraphics[width=1\textwidth]{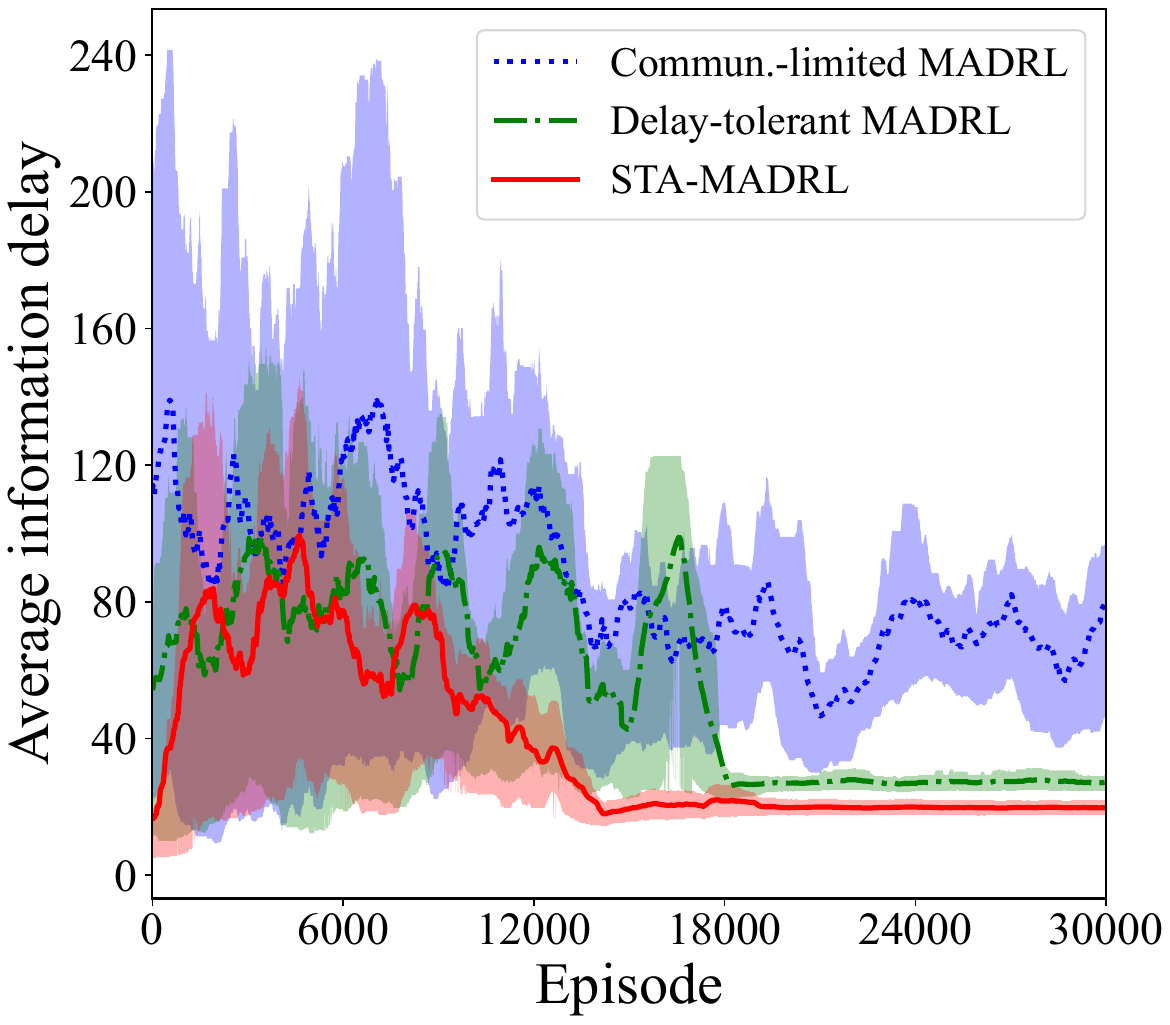}
		\end{minipage}
	}
	\caption{STA-MADRL achieves close throughput with the Ideal-MADRL.}
	\label{fig-conv-thu-delay}
	\vspace{-0.2cm}
\end{figure}

In Fig.~\ref{fig-conv-thu-delay}, we also compare the throughput and information delay performance of the STA-MADRL and the delay-tolerant MADRL algorithms, along with two baseline approaches. In our simulation, the STA-MADRL algorithm achieves 25\% throughput gain compared to the delay-tolerant MADRL and over 75\% throughput gain compared to the communication-limited MADRL at convergence. By correcting the delayed information using a spatio-temporal prediction module, the STA-MADRL algorithm offers more accurate awareness of the complete network environment, allowing UAVs to adapt their trajectories and transmission control strategies more effectively. Compared with Ideal-MADRL, the STA-MADRL algorithm achieves a very close throughput performance without real-time information sharing among all UAVs, making it more practical for deployment in UAV-assisted wireless networks. Fig.~\ref{fig-conv-thu-delay}(b) shows the UAVs' average information delay during the learning process. By penalizing the information delay, the delay-tolerant MADRL algorithm can significantly reduce the UAVs' average information delay, comparing with the communication-limited MADRL. The STA-MADRL algorithm further reduces the information delay by making more informative decisions based on the corrected network state. The average information delay can be reduced by 50\% with the communication-limited MADRL. Besides, we observe that STA-MADRL demonstrates a more stable learning performance and faster convergence, as both the dynamics of throughput and information delay in Fig.~\ref{fig-conv-thu-delay} show a smaller variance of fluctuations. The learning algorithms will converge as the variance of fluctuation stabilizes at a fixed level. By predicting the missing information, the STA-MADRL algorithm may provide smoother gradient updates during learning, leading to a more stable learning performance and faster convergence speed.

\begin{figure}[t]
	\centering
	\subfigure[Information delay over time.]{
		\begin{minipage}[b]{0.46\linewidth}
			\includegraphics[width=1\textwidth]{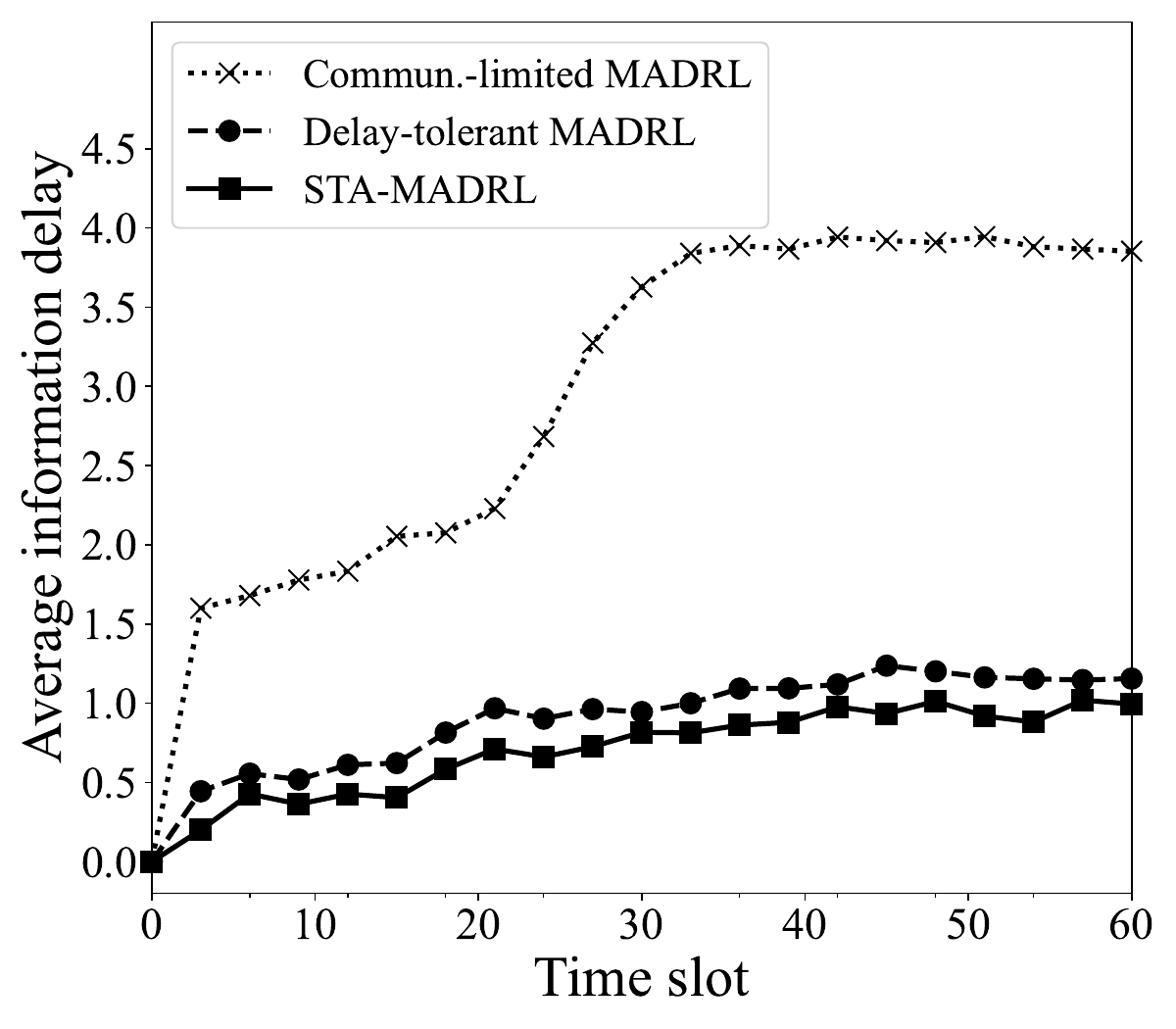}
		\end{minipage}
	}
	\subfigure[Variance of information delay.]{
		\begin{minipage}[b]{0.46\linewidth}
			\includegraphics[width=1\textwidth]{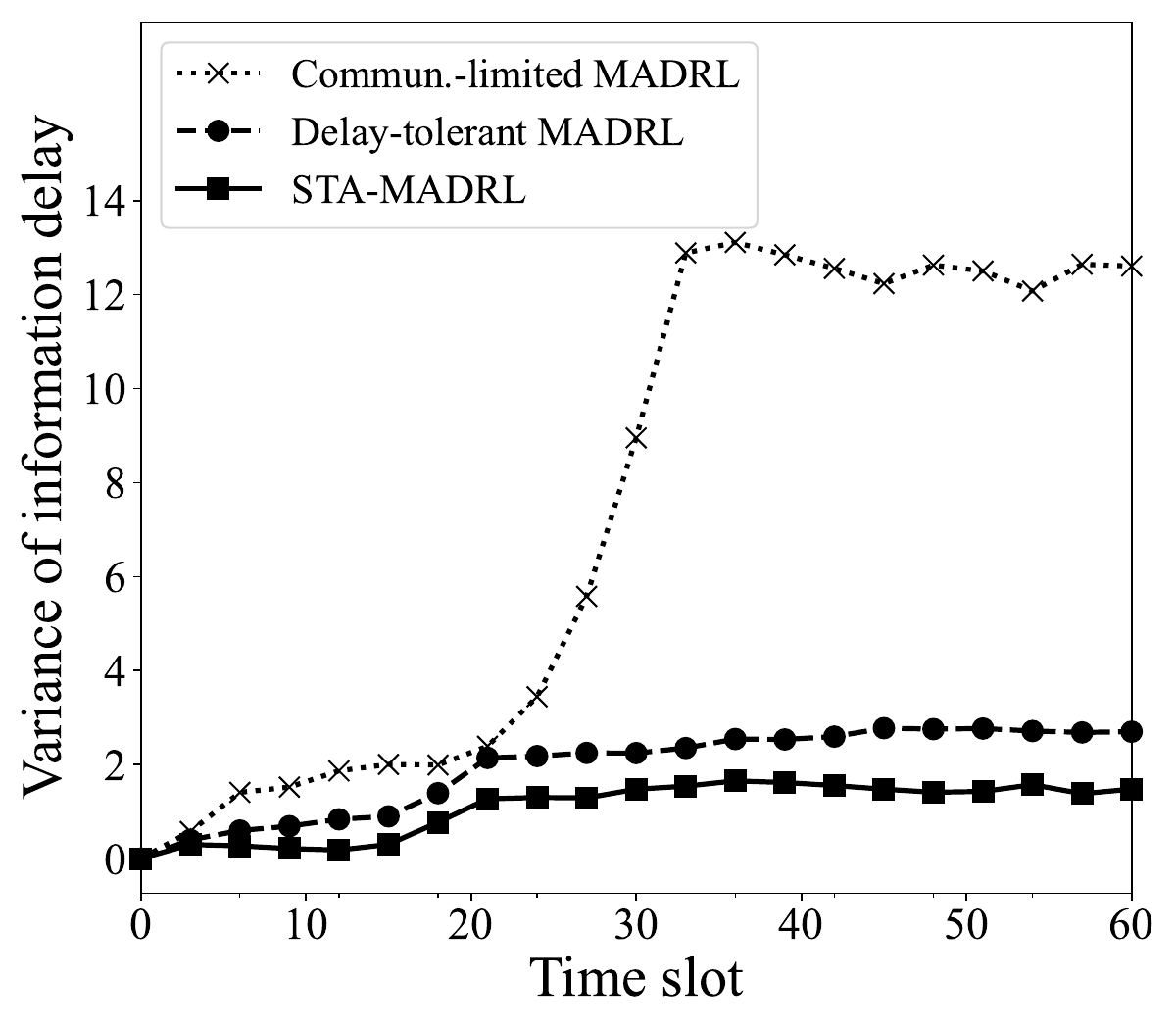}
		\end{minipage}
	}
	\caption{Delay-penalized reward improves fair information exchange.}
	\label{fig-delay-var}
\end{figure}

\begin{figure}[t]
	\centering
	\subfigure[U2B connection frequency.]{
		\begin{minipage}[b]{0.46\linewidth}
			\includegraphics[width=1\textwidth]{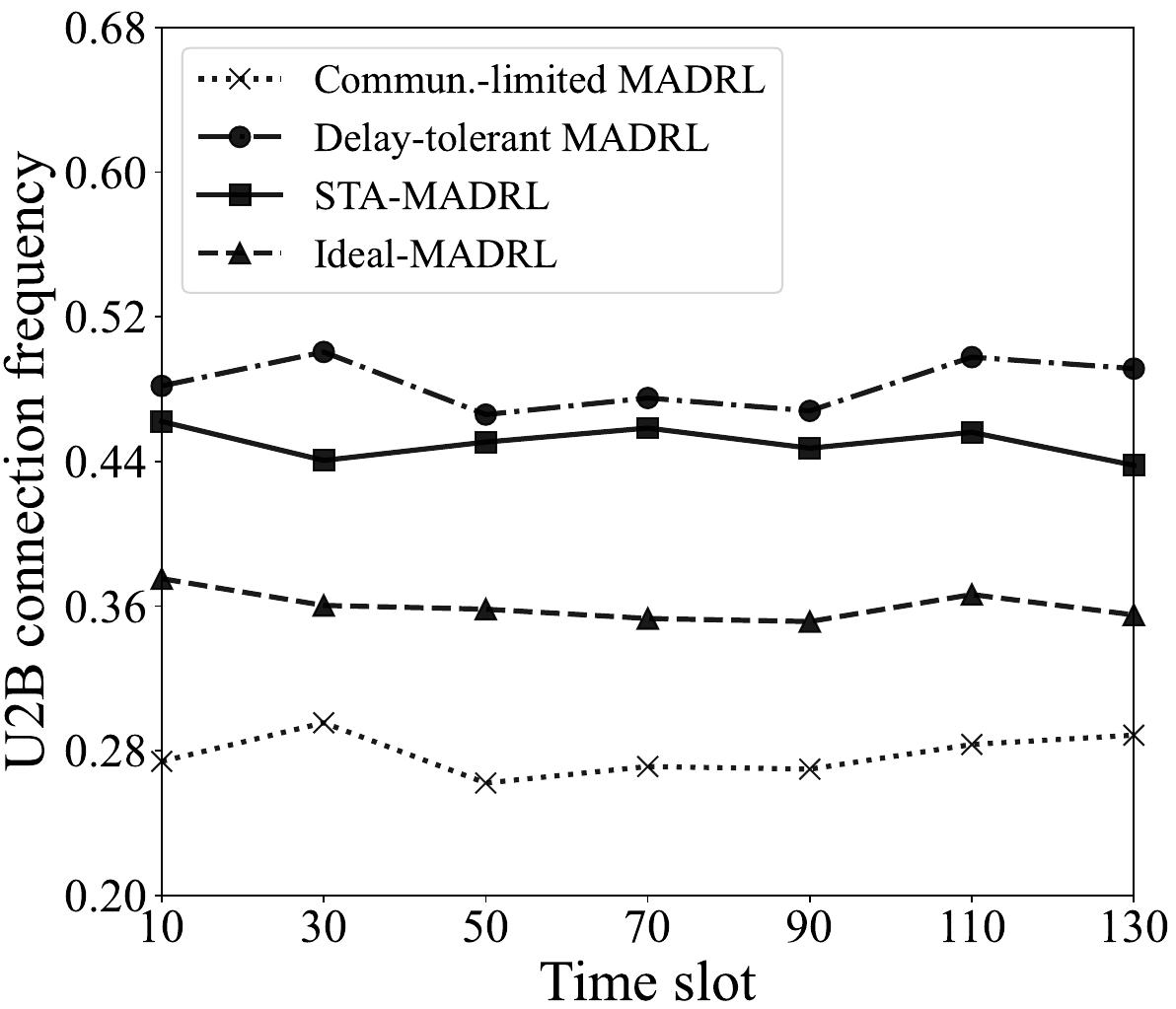}
		\end{minipage}
	}
	\subfigure[Information error.]{
		\begin{minipage}[b]{0.46\linewidth}
			\includegraphics[width=1\textwidth]{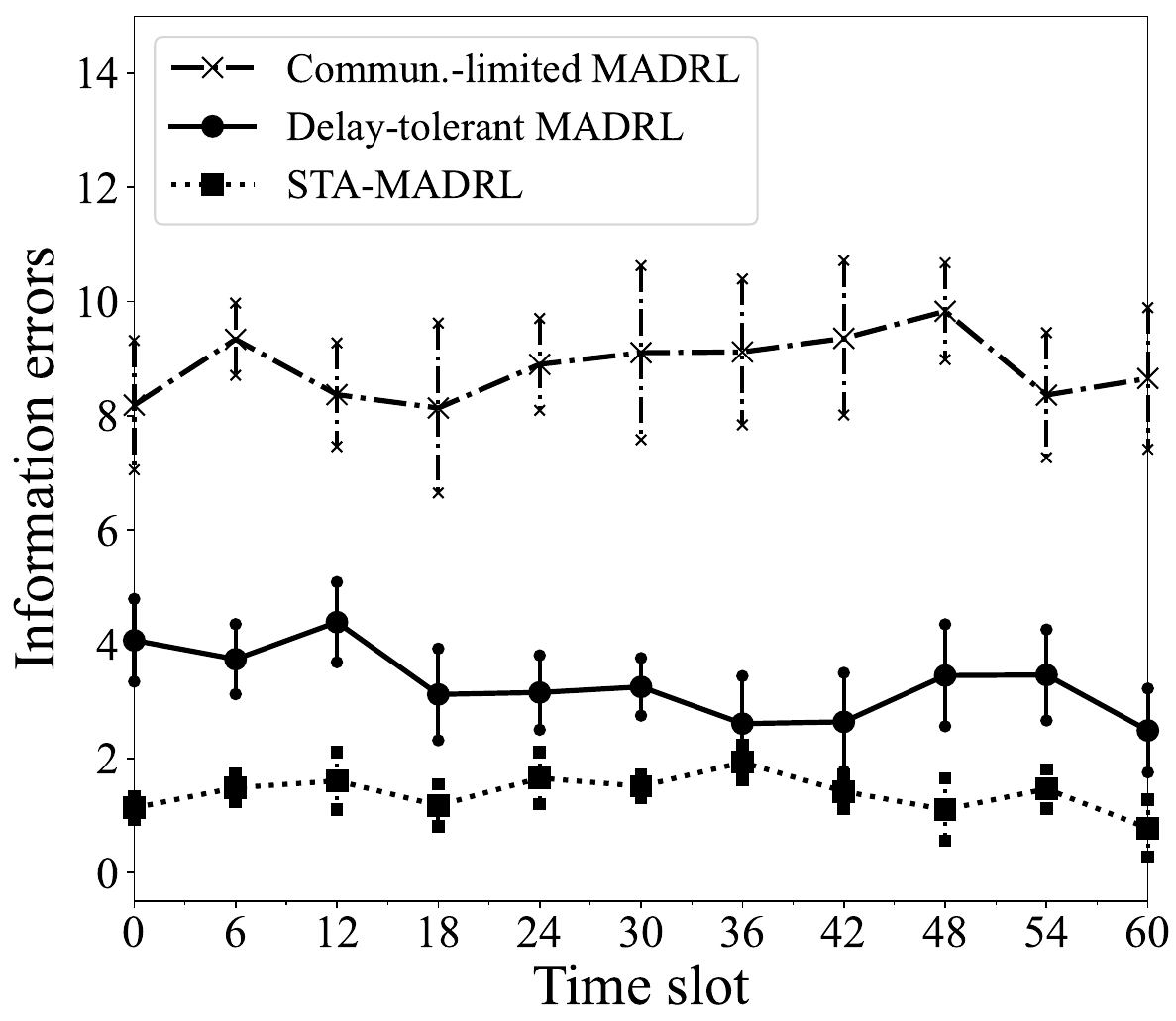}
		\end{minipage}
	}
	\caption{Frequent U2B information exchange reduces information error.}
	\label{fig-info-error}
	\vspace{-0.2cm}
\end{figure}

\subsection{Delay-penalized Reward Improves Information Sharing}
Figure~\ref{fig-delay-var} shows the UAVs' average information delay and delay variance over time along their trajectories. Compared with the communication-limited MADRL algorithm, the delay-penalized reward encourages the UAVs to have more frequent information exchange with the BS and thus it can maintain the UAVs' information delay at a relatively low level as shown in Fig.~\ref{fig-delay-var}(a). A smaller variance of the UAVs' information delay in Fig.~\ref{fig-delay-var}(b) implies that all UAVs in the STA-MADRL and the delay-tolerant MADRL algorithms can fairly have the opportunities to exchange state information with the BS. This verifies that the delay-penalized reward design can incentivize UAVs to maintain well-balanced information delay at a low level. Without such a delay incentive, the communication-limited MADRL algorithm may allocate one UAV to serve the remote GUs for a long time period, leading to both a higher average delay in Fig.~\ref{fig-delay-var}(a) and the unbalanced delay among UAVs in Fig.~\ref{fig-delay-var}(b).


\begin{figure*}[t]
\centering
\subfigure[Ideal-MADRL]{
	\begin{minipage}[b]{0.23\linewidth}
	\includegraphics[width=1\textwidth]{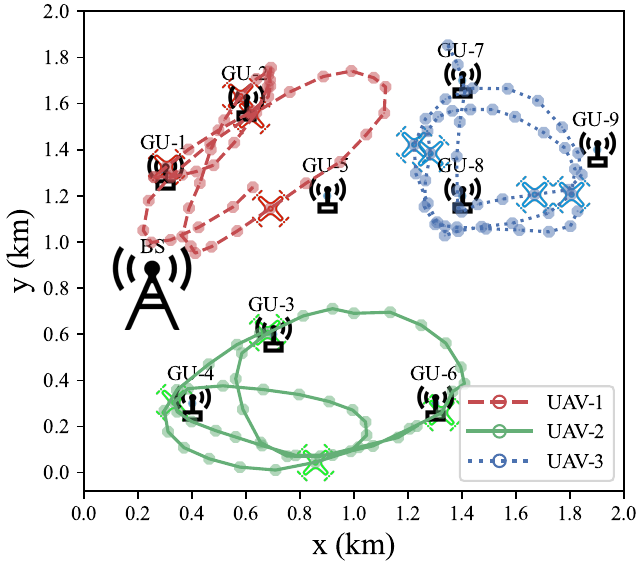}
	\end{minipage}
	}
\subfigure[Communication-limited MADRL]{
  	\begin{minipage}[b]{0.23\linewidth}
   	\includegraphics[width=1\textwidth]{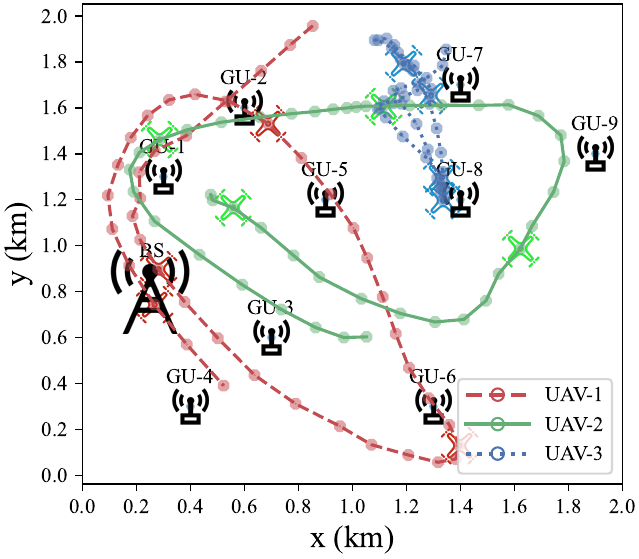}
  	\end{minipage}
   	}
\subfigure[Delay-tolerant MADRL]{
	\begin{minipage}[b]{0.23\linewidth}
 	\includegraphics[width=1\textwidth]{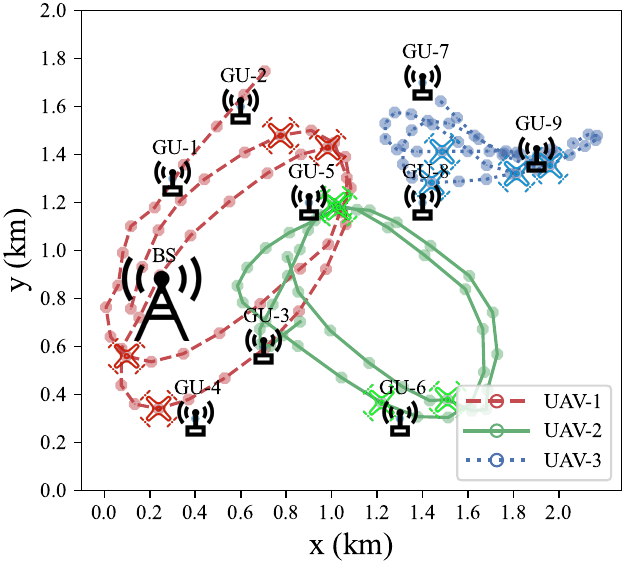}
   	\end{minipage}
   	}
\subfigure[STA-MADRL]{
	\begin{minipage}[b]{0.23\linewidth}
	\includegraphics[width=1\textwidth]{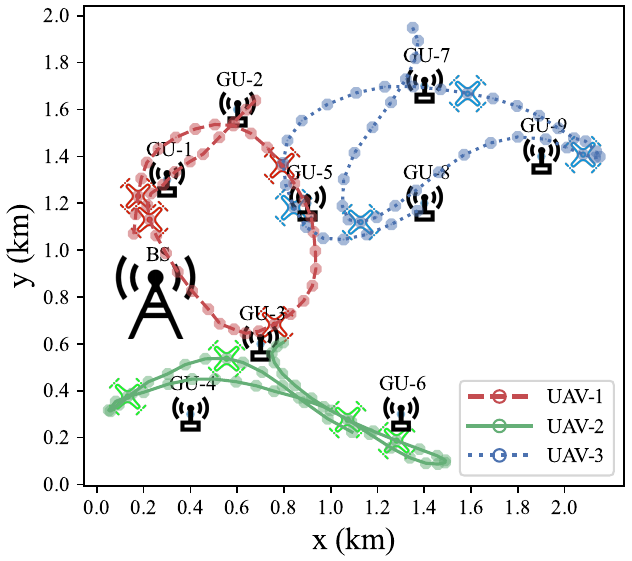}
	\end{minipage}
    }
\caption{The UAVs' trajectory planning with different learning algorithms.}\label{fig-4trajectory}
\vspace{-0.2cm}
\end{figure*}

Figure~\ref{fig-info-error}(a) shows the UAVs' average frequencies of information exchange with the BS in different algorithms, which are obtained by counting and averaging the number of the UAVs' U2B connections on their trajectories over different time slots. Clearly, relying on the delay-penalized reward, the delay-tolerant MADRL and STA-MADRL algorithms achieve more frequent U2B connections than that of the communication-limited MADRL. As such, the BS can timely update all UAVs' state information and share the complete state with other UAVs, enhancing the UAVs' environmental awareness and promoting their multi-agent collaborations. An interesting observation is that the STA-MADRL can tolerate a slightly smaller U2B connection frequency compared to the delay-tolerant MADRL algorithm. This is because the spatio-temporal prediction module in the STA-MADRL can help UAVs to recover the delayed information and build the complete system state without frequent U2B connections. As such, the UAVs in the STA-MADRL algorithm can have more opportunities to serve the GUs and thus improve the network capacity, as revealed in Fig.~\ref{fig-conv-thu-delay}(a). The reward performance of the communication-limited MADRL degrades significantly due to the misaligned information input $(\bm o_n(t),\tilde{\bm{o}}_{-n}(t))$ to the UAV-$n$'s actor network, where $\tilde{\bm{o}}_{-n}(t)$ is the delayed information from the other UAVs. The STA-MADRL algorithm actually intends to minimize the information error between $\tilde{\bm{o}}_{-n}(t)$ and the real-time state ${\bm{o}}_{-n}(t)$ by using the spatio-temporal attention based prediction module. In Fig.~\ref{fig-info-error}(b), we characterize each UAV's information error by $|\tilde{\bm{o}}_{-n}(t)-{\bm{o}}_{-n}(t)|$ and compare the UAVs' average information error in different learning algorithms. It is clear that the STA-MADRL algorithm maintains the lowest information error, which provides a more accurate estimate of the complete system state for the UAVs' decision-making in the MADRL framework. The comparison results in Fig.~\ref{fig-info-error}(b) corroborates with the observations in Fig.~\ref{fig-info-error}(a). Taking communication-limited MADRL as the baseline, a higher U2B connection frequency with the delay-tolerant MADRL algorithm in Fig.~\ref{fig-info-error}(a) implies less information error in Fig.~\ref{fig-info-error}(b).

\subsection{Information Sharing Enhances Network Throughput}

The delay-penalized reward design in~\eqref{equ-reward-delay} cares about both the UAVs' transmission performance and the opportunities for information sharing by planning the UAVs' trajectories. To ensure frequent information exchange with the BS, a distant UAV has to circulate back to report its local information to the BS. Intuitively this may sacrifice the UAVs' network throughput to serve the GUs, i.e.,~the UAVs' transmission performance may degrade when they fly back and forth between the BS and their service areas. However, our counter-intuitive observation is that the UAVs' throughput performance can also be increased significantly by planning frequent U2B connections on the UAVs' trajectories. As shown in Fig.~\ref{fig-conv-thu-delay}(a), the convergent throughput performance in STA-MADRL is close to the optimum achievable by the Ideal-MADRL and nearly doubled compared to that of the communication-limited MADRL. This is because the frequent information sharing provides accurate estimation of the complete system state and supports more efficient collaboration among UAVs to improve the overall network throughput.

In this part, we further evaluate the UAVs' throughput on their trajectories with different algorithms. Fig.~\ref{fig-4trajectory} shows the UAVs' trajectories in the STA-MADRL and delay-tolerant MADRL algorithms, as well as two baselines, to serve the GUs under the same setting. The UAVs' trajectories are highlighted with different line styles. In Fig.~\ref{fig-4trajectory}(a), Ideal-MADRL achieves full network coverage with balanced service of all GUs. With complete information, each UAV can be optimally allocated to serve a subset of GUs with a clear boundary and stable trajectory. In contrast, the communication-limited MADRL algorithm in Fig.~\ref{fig-4trajectory}(b) leads to conflicting trajectories with obvious overlaps on individual UAVs' service areas. Besides service overlap, some UAVs are assigned with excessively large service areas, e.g.,~the red and green UAVs. This may lead to excessive information delay when they are far from the BS for a long time. Without timely information exchange, each UAV will try to cover the whole service area and serve all GUs, as shown in Fig.~\ref{fig-4trajectory}(b). This explains how excessive information delay hinders the UAV's efficient collaboration. In Fig.~\ref{fig-4trajectory}(c) and Fig.~\ref{fig-4trajectory}(d), though the delay-tolerant MADRL and STA-MADRL produce different trajectories, one common observation is that all UAVs have relatively stable routes and serve a subset of GUs with little overlap, compared to the communication-limited MADRL. The delay-penalized rewards in both algorithms drive the UAVs to periodically circulate above a subset of GUs. Compared with STA-MADRL, the delay-tolerant MADRL still has some overlapping service areas and some uncertainties in the UAVs' trajectories, due to the lack of real-time complete system state. Compared with Ideal-MADRL, STA-MADRL also performs well, i.e.,~each UAV's trajectory has a clear boundary with minimum overlap on the service area.

\begin{figure}[t]
	\centering
	\subfigure[Accumulated throughput]{
		\begin{minipage}[b]{0.44\linewidth}
			\includegraphics[width=1\textwidth]{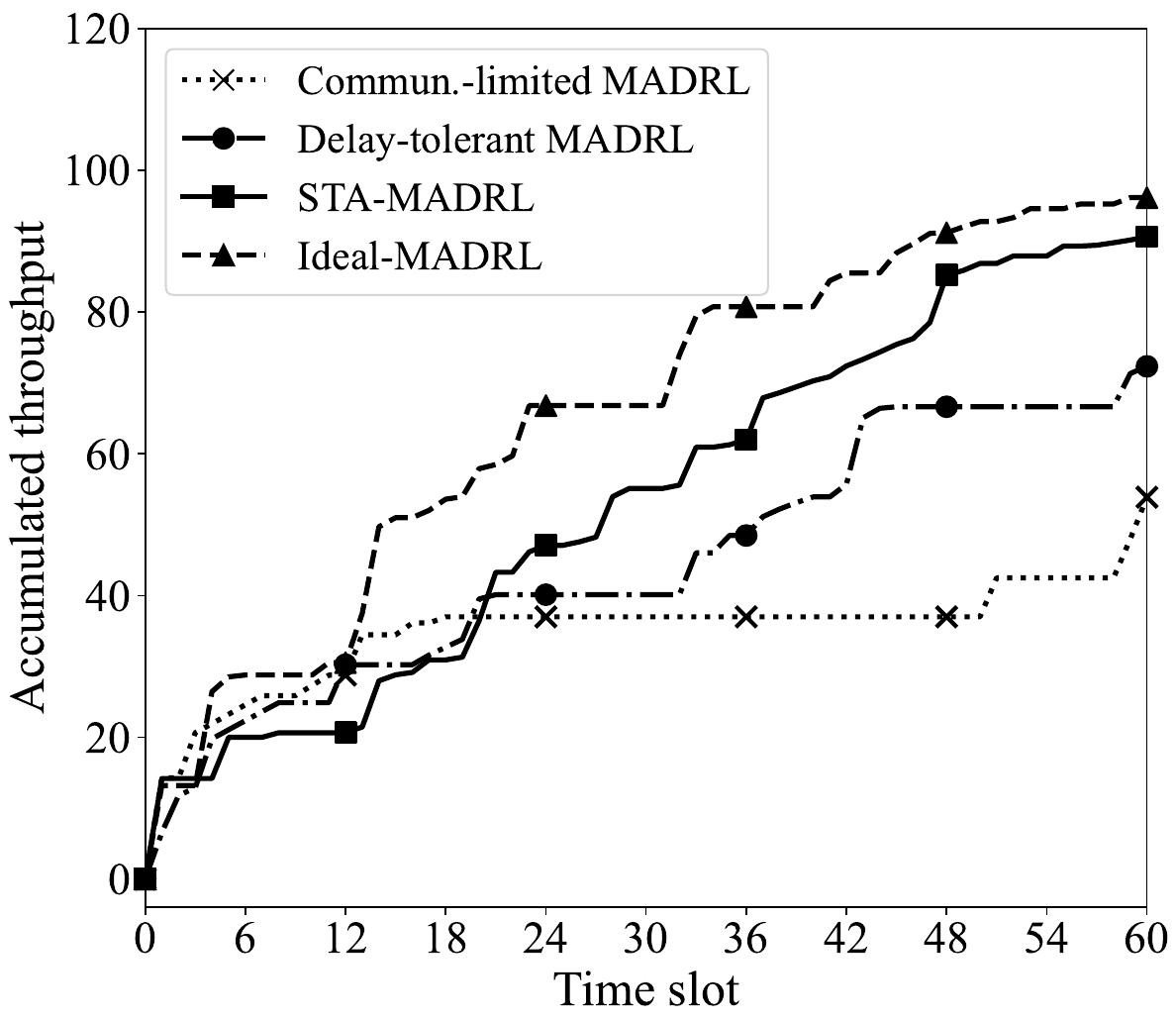}
		\end{minipage}
	}
	\subfigure[STA-MADRL's max. throughput]{
		\begin{minipage}[b]{0.47\linewidth}
			\includegraphics[width=1\textwidth]{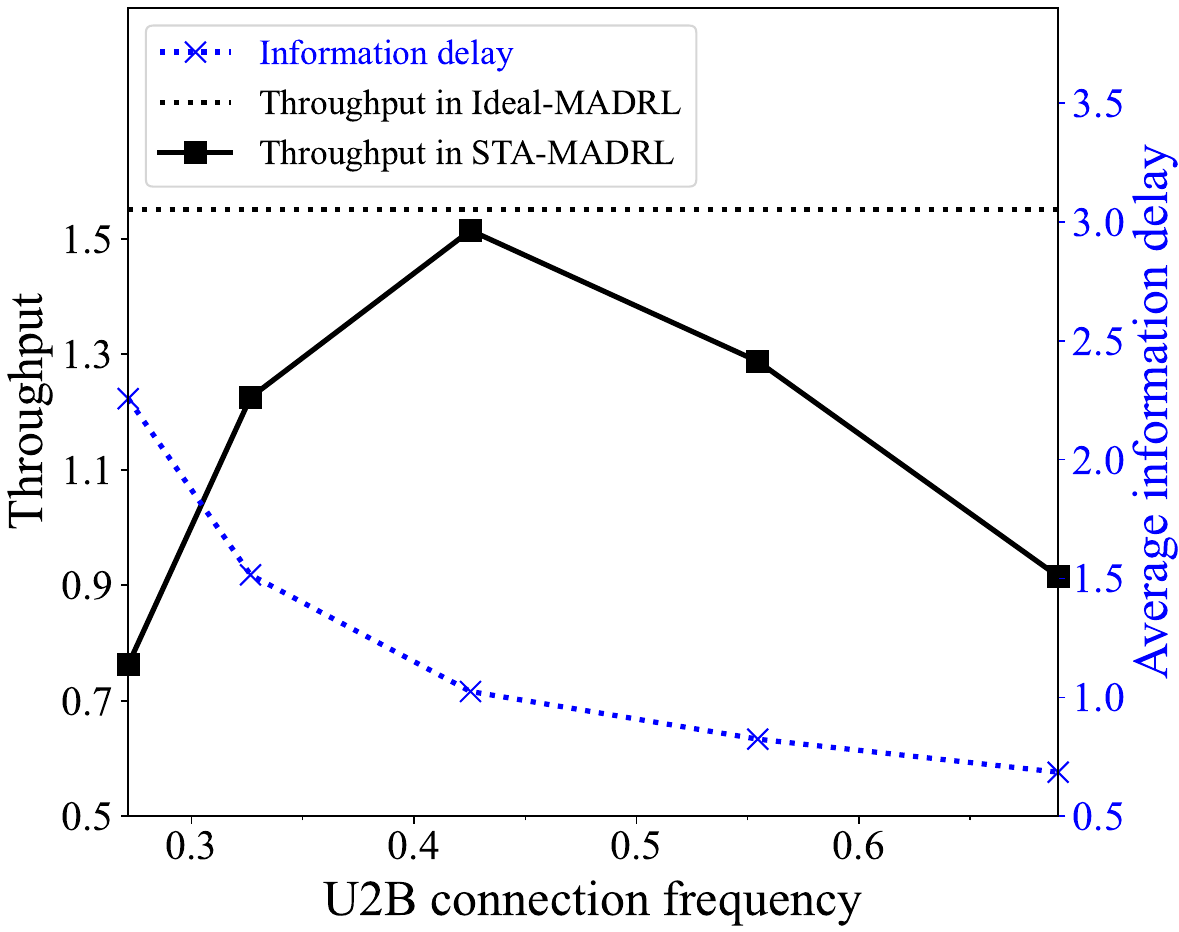}
		\end{minipage}
	}
	\caption{Throughput performance with different learning algorithms.}
	\label{fig-network-thu}
\end{figure}
\begin{figure}[t]
	\centering
	\subfigure[STA-MADRL]{
		\begin{minipage}[b]{0.46\linewidth}
			\includegraphics[width=1\textwidth]{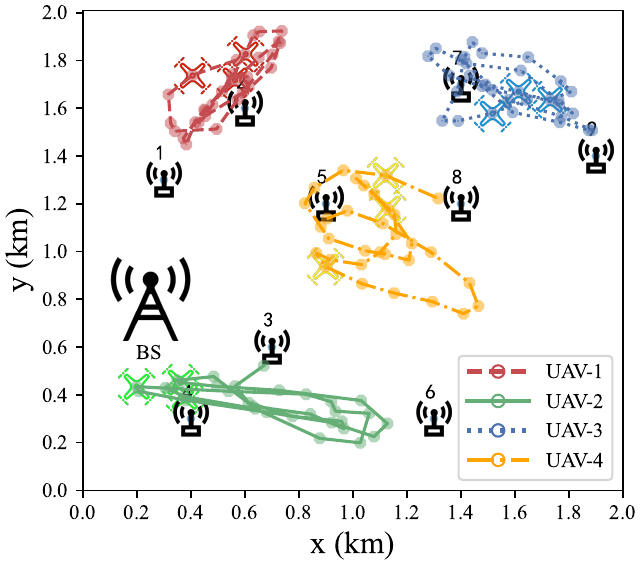}
		\end{minipage}
	}
	\subfigure[Commun.-limited  MADRL]{
		\begin{minipage}[b]{0.46\linewidth}
			\includegraphics[width=1\textwidth]{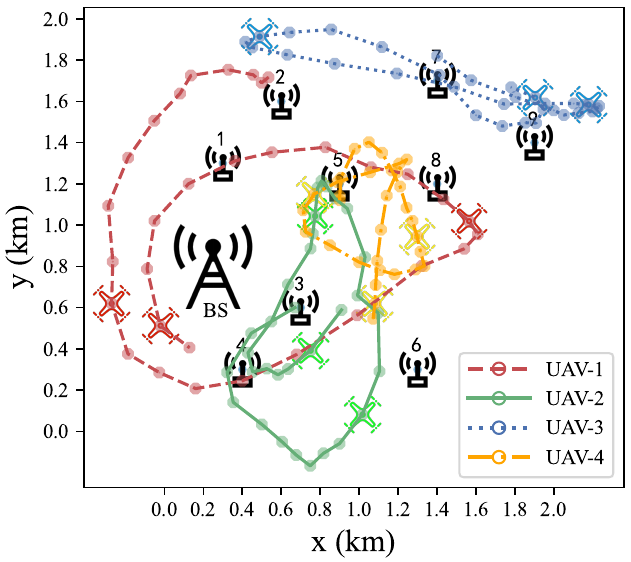}
		\end{minipage}
	}
	\caption{Clear boundary with less service overlap in STA-MADRL.}
	\label{fig-4UAV-traj}
	\vspace{-0.2cm}
\end{figure}


\subsection{Trade-off between Learning and Throughput Performance}

Correspondingly, Fig.~\ref{fig-network-thu}(a) records the BS's accumulated throughput as the UAVs move on their trajectories. With complete information, the Ideal-MADRL algorithm achieves the highest throughput in Fig.~\ref{fig-network-thu}(a) and serves as the upper bound. The service overlaps in the communication-limited MADRL algorithm inevitably decrease the efficiency of UAVs' collaboration and result in a significant throughput degradation, which can be regarded as the throughput lower bound for our algorithm design. The accumulated throughput with STA-MADRL is continuously increasing and close to that of Ideal-MADRL as shown in Fig.~\ref{fig-network-thu}(a). The above results reveal that the network throughput and learning performance can be improved significantly and simultaneously by enabling regularly frequent information sharing among UAVs.

However, we also envision that the cost of information exchange cannot be ignored when excessive channel resources are consumed to establish real-time U2B connections. In an extreme case, all UAVs will stay close to the BS and exchange real-time information with the BS. Such real-time information exchange is useful for the awareness of network environment and multi-agent decision making in the MADRL framework. However, it inevitably limits the network coverage and capacity. In this part, we examine how the overall network throughput is affected by the U2B connection frequency. Fig.~\ref{fig-network-thu}(b) shows the change of throughput with respect to the UAVs' average U2B connection frequencies. Given the weighting coefficient $\omega_2$ in the delay-penalized reward~\eqref{equ-reward-delay}, we actually fix the UAVs' sensitivities to information delay and will observe the U2B connection frequency in Fig.~\ref{fig-info-error} by planning trajectories with the STA-MADRL algorithm. Hence, the $x$-axis in Fig.~\ref{fig-network-thu}(b) is obtained by varying the weighting coefficient $\omega_2$, while the $y$-axes denote the corresponding throughput and delay performance achieved by the STA-MADRL algorithm, respectively.

\begin{figure}[t]
	\centering
	\includegraphics[width=0.9\linewidth]{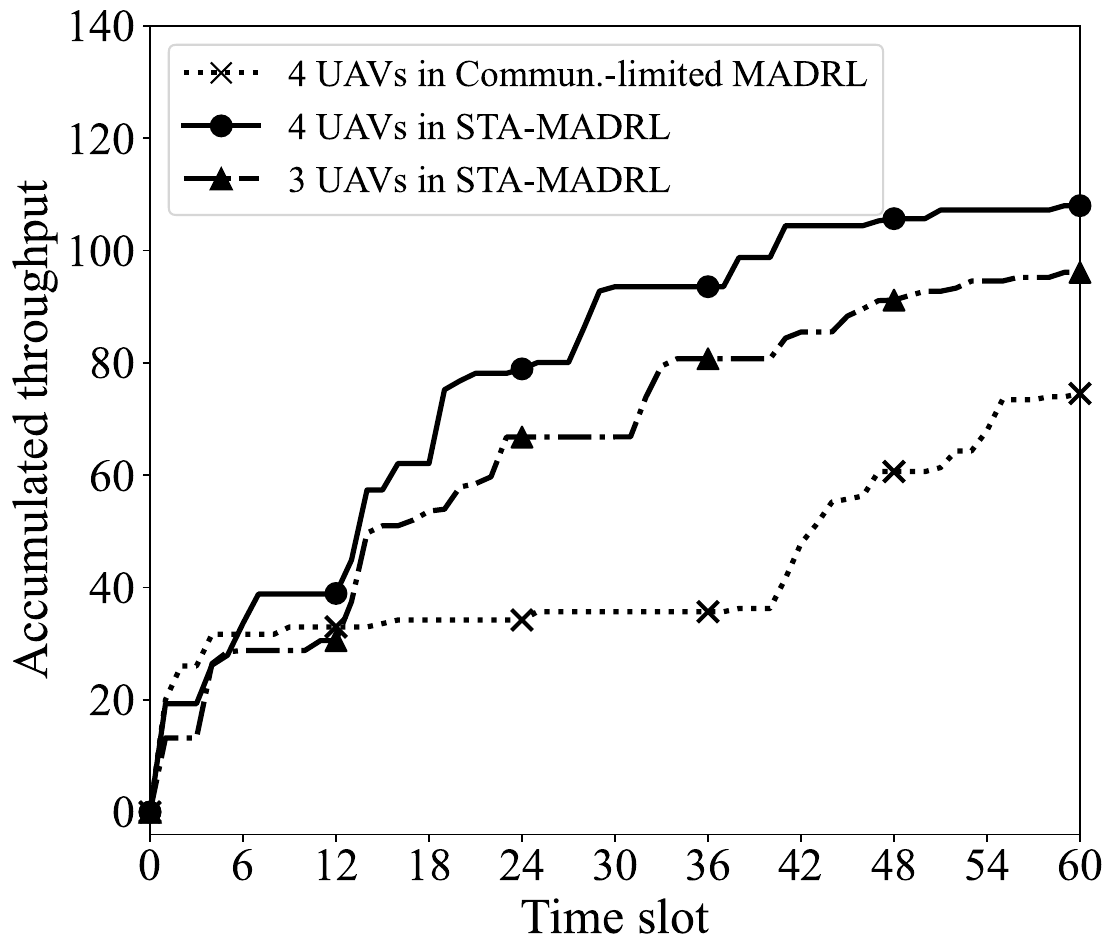}
	\caption{STA-MADRL achieves a higher throughput with a fewer UAVs.}
	\label{fig-robust-thu}
	\vspace{-0.3cm}
\end{figure}

An interesting observation is that the network throughput with STA-MADRL is not always increasing with the UAVs' frequencies of information exchange. As shown in Fig.~\ref{fig-network-thu}(b), the network throughput first increases with the U2B connection frequency and then declines as we further increase the U2B connection frequency. There is a clear optimal frequency of information exchange that achieves the maximum throughput, which is close to that of Ideal-MADRL. It implies that real-time information exchange is not necessary for a practical UAV-assisted wireless network, considering the trade-off between throughput and learning performance. With low U2B connection frequencies, the UAVs experience severe information delay and estimation error in the complete system state, leading to throughput degradation in the UAVs' collaborative control. Conversely, when the UAVs become more sensitive to information delays and demand excessive information exchange, substantial channel resources will be consumed by information exchange, which restricts the UAVs' transmission capabilities to forward the GUs' sensing data.

\subsection{Robustness of Multi-UAV's Collaborative Control}

The delay-penalized reward design constantly monitors each UAV's information delay and forces frequent information exchange with the BS, while the spatio-temporal attention based prediction module further recovers the missing information. These will help each UAV estimate the real-time complete system state and adapt to the network environment. Therefore, we expect that STA-MADRL will be more robust against the network dynamics in optimizing the UAVs' trajectories and transmission control strategies. Compared with the 3-UAVs' trajectory planning in Fig.~\ref{fig-4trajectory}, we add a new UAV into the service area and show the new trajectory planning strategy for the current 4 UAVs scenario in Fig.~\ref{fig-4UAV-traj}. By enabling frequent information exchange and prediction, STA-MADRL can detect the presence of a new UAV and optimize its trajectory jointly with other UAVs. Thus, we observe that STA-MADRL can adaptively adjust each UAV's service area to ensure a clear boundary with minimum service overlap, as shown in Fig.~\ref{fig-4UAV-traj}(a). In contrast, with the same network settings, the UAVs' trajectories in the communication-limited MADRL become quite chaotic, as shown in Fig.~\ref{fig-4UAV-traj}(b), due to a lack of accurate state information of the network. Such a lack of information sharing evidently limits the efficiency of UAVs' collaborative control, and also makes it inflexible and insensitive to the change of network environment.

To evaluate the efficiency of collaboration, we quantitatively compare the throughput accumulated at the BS as 4 UAVs fly along the trajectories in Fig.~\ref{fig-robust-thu}. Without doubt STA-MADRL achieves over a 40\% throughput gain over the communication-limited MADRL algorithm. Besides, we plot STA-MADRL's throughput with 3 UAVs serving the same set of GUs in Fig.~\ref{fig-robust-thu}. An interesting observation is that 3 UAVs with STA-MADRL can even achieve a higher throughput than that of 4 UAVs with the communication-limited MADRL algorithm. This observation further verifies that STA-MADRL is robust to the network dynamics and more efficient for UAVs' collaborative control.

\begin{figure}[t]
	\centering
	\subfigure[Throughput]{
		\begin{minipage}[b]{0.45\linewidth}
			\includegraphics[width=1\textwidth]{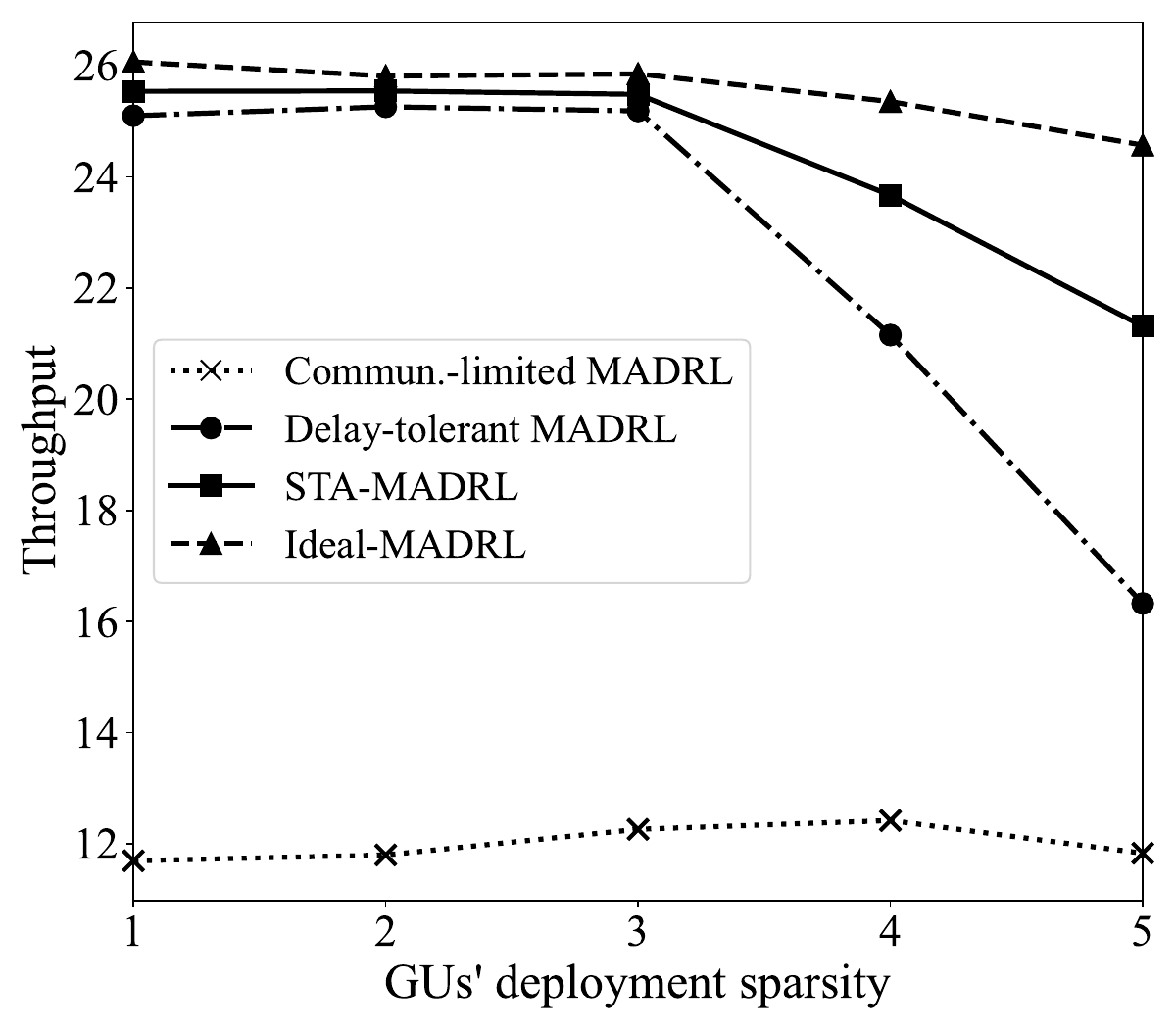}
		\end{minipage}
	} \
	\subfigure[Information delay]{
		\begin{minipage}[b]{0.45\linewidth}
			\includegraphics[width=1\textwidth]{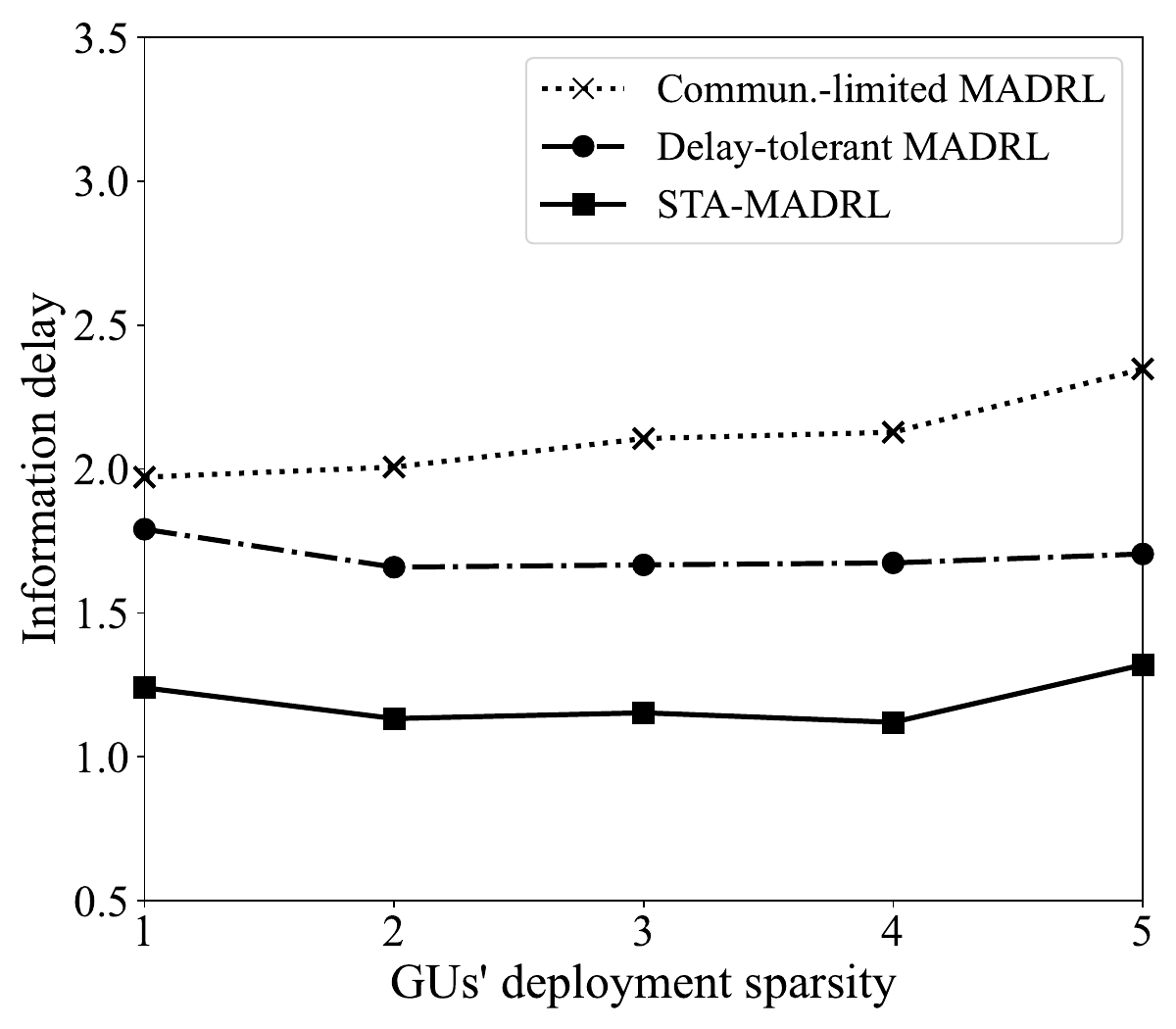}
		\end{minipage}
	}
	\caption{The impact of GUs' deployment sparsity.}
	\label{fig-density}
	\vspace{-0.2cm}
\end{figure}

Figure~\ref{fig-density} shows the impact of GUs' deployment sparsity on throughput and information delay, where the sparsity level increases from 1 to 5. As depicted in Fig.~\ref{fig-density}(a), STA-MADRL achieves throughput comparable to Ideal-MADRL and outperforms communication-limited MADRL across all sparsity levels. While throughput decreases for most schemes as deployment becomes sparser due to deteriorated channel conditions, STA-MADRL maintains relatively stable performance. Fig.~\ref{fig-density}(b) demonstrates that STA-MADRL effectively reduces information delay compared to delay-tolerant MADRL, which validates the effectiveness of spatio-temporal prediction in compensating for communication delays.

\vspace{-0.2cm}
\section{Conclusions} \label{sec-con}
In this paper, we have focused on trajectory planning, network formation, and transmission control, in a multi-UAV-assisted wireless network with limited communications. We have addressed the performance degradation in the conventional communication-limited MADRL framework by two special designs. The delay-penalized reward firstly encourages each UAV to plan a proper trajectory that supports frequent information exchange with the BS. The spatio-temporal attention module exploits the UAVs' historical information for an enhanced awareness of complete network state and more efficient collaborative control in the MADRL framework. Numerical results showed that the proposed STA-MADRL achieves more favorable delay and throughput performance. It also verified the trade-off between learning and throughput performance. The network throughput and learning performance can be improved significantly and simultaneously with a proper frequency of information exchange.


\bibliographystyle{IEEEtran}
\bibliography{limited-comm-uav}
\end{document}